\documentclass[reprint,amsmath,amssymb,aps,pre,superscriptaddress,longbibliography]{revtex4-1}
\selectfont

\usepackage[utf8]{inputenc}
\usepackage[dvipsnames]{xcolor}
\usepackage[colorlinks=true,linkcolor=Blue,citecolor=Blue,urlcolor=Blue]{hyperref}
\usepackage{graphicx}
\usepackage{subfigure}

\newcommand{\beq}{\begin{equation}}
\newcommand{\eeq}{\end{equation}}
\newcommand{\bnab}{\boldsymbol{\nabla}}
\newcommand{\ed}{\mathrm{d}}

\begin{document}

\title{Wada structures in a binary black hole system}

\author{\'{A}lvar Daza}
\email{alvar.daza@urjc.es}
\affiliation{Nonlinear Dynamics, Chaos and Complex Systems Group, Departamento de F\'{i}sica, Universidad Rey Juan Carlos, M\'{o}stoles, Madrid, Tulip\'{a}n s/n, 28933, Spain}

\author{Jake O.~Shipley}
\email{joshipley1@sheffield.ac.uk}
\affiliation{Consortium for Fundamental Physics, School of Mathematics and Statistics, University of Sheffield, Hicks Building, Hounsfield Road, Sheffield S3 7RH, United Kingdom}

\author{Sam R.~Dolan}
\email{s.dolan@sheffield.ac.uk}
\affiliation{Consortium for Fundamental Physics, School of Mathematics and Statistics, University of Sheffield, Hicks Building, Hounsfield Road, Sheffield S3 7RH, United Kingdom}

\author{Miguel A.~F.~Sanju\'{a}n}
\email{miguel.sanjuan@urjc.es}
\affiliation{Nonlinear Dynamics, Chaos and Complex Systems Group, Departamento de F\'{i}sica, Universidad Rey Juan Carlos, M\'{o}stoles, Madrid, Tulip\'{a}n s/n, 28933, Spain}
\affiliation{Department of Applied Informatics, Kaunas University of Technology, Studentu 50-415, Kaunas LT-51368, Lithuania}
\affiliation{Institute for
Physical Science and Technology, University of Maryland, College
Park, Maryland 20742, USA}

\date{\today}

\begin{abstract}
A key goal of the Event Horizon Telescope is to observe the shadow cast by a black hole. Recent simulations have shown that binary black holes, the progenitors of gravitational waves, present shadows with fractal structure. Here we study the binary shadow structure using techniques from nonlinear dynamics, recognising shadows as exit basins of open Hamiltonian dynamical systems. We apply a recently developed numerical algorithm to demonstrate that parts of the Majumdar--Papapetrou binary black hole shadow exhibit the Wada property: any point of the boundary of one basin is also on the boundary of at least two additional basins. We show that the algorithm successfully distinguishes between the fractal and regular (i.e., non-fractal) parts of the binary shadow.
\end{abstract}

\maketitle

\section{Introduction \label{sec:intro}}

Einstein's general theory of relativity and chaos theory/nonlinear dynamics are two of the deepest conceptual advances of twentieth-century science. The former changed our perception of space, time and gravity, and the latter showed how deterministic rules give rise to chaotic behaviour if nonlinearities are involved. General relativity -- itself a nonlinear field theory -- naturally leads to deterministic chaos. For example, the fate of a photon approaching a pair of black holes (BHs) can be essentially indeterminate (we shall show), even though it is governed by a deterministic set of equations. In this article we explore a topic of interest to astronomers, relativists, and nonlinear dynamicists alike: the intricate structure of the shadow cast by the event horizons of a pair of BHs.

An exciting era for gravitational astronomy is underway. In 2015, the first direct observation of gravitational waves (GWs) \cite{AbbottEtAl2016}, by the LIGO/Virgo collaboration, confirmed that binary BHs exist in Nature. In 2017, a GW signal from a binary neutron star inspiral was accompanied, $\sim 1.7~\text{s}$ later, by a gamma-ray burst \cite{AbbottEtAl2017}. The Event Horizon Telescope (EHT) -- a  world-scale telescopic array employing millimeter-wavelength very-long-baseline interferometry (VLBI) \cite{FalckeMeliaAgol2000} -- has begun observing nearby galactic centres. The goal of the EHT is to image the environment of astrophysical BH candidates, such as Sagittarius A$^{\ast}$.  A key target of the EHT is to resolve the \emph{shadow} cast by the event horizon of a supermassive BH itself \cite{BroderickJohannsenLoebEtAl2014}. High-resolution images from the EHT will enable the first tests of the no-hair conjecture \cite{BroderickJohannsenLoebEtAl2014}, which asserts that BHs are characterised by just three quantities: mass, angular momentum, and charge (with the latter thought to be negligible).

A feature of Einstein's theory is the \emph{gravitational lensing} of light \cite{Perlick2004}. Massive bodies, such as stars or BHs, generate spacetime curvature, leading to the deviation in the trajectories of photons as they trace out null geodesics (``rays'') on the curved geometry. A BH shadow is a region in the observer's sky which cannot be illuminated by distant light sources, due to the blockage of the BH. Equivalently, the shadow is associated with the set of all photons which, when traced backwards in time from the observer, asymptote towards the \emph{event horizon} of the BH. For a recent review of BH shadows and strong-field gravitational lensing, see Ref.~\cite{CunhaHerdeiro2018}.

In the language of nonlinear dynamics, a BH shadow is an exit basin \cite{nusse_basins_1996,Aguirre2001} in an open Hamiltonian dynamical system. The exit basin is defined on initial-data surface for null geodesics; typically this is taken to be the image plane of a distant observer. The BH shadow is the set of initial conditions that lead to a particular region of phase space, namely, the event horizon of the BH.

Motivated by the GW detections from merging binary BHs, and the future prospects of the EHT, a strand of recent work has focussed on what the shadow of a \emph{pair} of BHs would look like, both for realistic dynamical binaries \cite{BohnThroweHebertEtAl2015}, and imitative models \cite{NittaChibaSugiyama2011,YumotoNittaChibaEtAl2012,ShipleyDolan2016,CunhaHerdeiroRodriguez2018}. In the former case, the lensing phenomena and BH shadows are studied using full nonlinear numerical simulations of the field equations; this a computationally expensive and technically demanding exercise. To build a qualitative understanding of binary shadows, one may instead study exact solutions with additional symmetries (e.g.~stationarity or axisymmetry), such as the Majumdar--Papapetrou binary BH in which two extremally charged BHs are in static equilibrium. Imitative models have been shown to capture some of the lensing phenomena associated with dynamical binary BHs (see e.g.~\cite{CunhaHerdeiroRodriguez2018}).

The presence of a pair of BHs reduces the symmetry (formally, by eliminating the Killing tensor associated with the Carter constant). As a result, the null geodesic equations, which describe the propagation of photons, are non-integrable, and chaotic scattering of photons emerges naturally. One of the hallmarks of chaos is the presence of fractal structures in phase space \cite{Aguirre2009}. For a discussion of fractal structures in the MP binary BH system, see \cite{Contopoulos1990, Contopoulos1991, Contopoulos2002, DettmannFrankelCornish1994, DettmannFrankelCornish1995, CornishGibbons1997, ShipleyDolan2016, DolanShipley2016}.

In a binary BH system, a photon meets one of three possible fates: it falls into the first BH, the second  BH, or it escapes to infinity. Thus it is natural to define three exit basins. As we shall show, across the phase space the shadow may exhibit both a regular (i.e., non-fractal) and a fractal structure. Furthermore, in certain parts of the phase space, the three basins have the more restrictive property of Wada, with all three basins sharing a common fractal boundary.

Just over one hundred years ago, the \emph{lakes of Wada} were proposed by Yoneyama \cite{yoneyama_theory_1917} as a curious example of three open sets in the plane which all share the same fractal boundary. In 1991, Kennedy and Yorke \cite{KennedyYorke1991} showed that open sets with this intriguing property are not only a topological curiosity, but they also occur in dynamical systems. Since then, the Wada property has been found in the basins of a range of chaotic dynamical systems, including the Gaspard--Rice three-disc system, the H\'{e}non--Heiles Hamiltonian and the Duffing oscillator (see e.g.~\cite{Aguirre2009}).

One of the main consequences of the existence of Wada basins in phase space is the difficulty which arises when predicting the final state of a particle. If there are small uncertainties in fixing initial conditions close to a Wada boundary, one encounters a high level of indeterminacy and an extreme sensitive dependence on initial conditions, despite the system being fully deterministic. For the binary BH system, this means that a photon which starts close to a Wada boundary in phase space could end up in one of three final states: the photon could fall into either of the BHs, or escape to spatial infinity.

Here we apply a recently developed numerical method \cite{daza_ascertaining_2018} to test for the Wada property, based on merging basins together in a pairwise fashion. In Ref.~\cite{daza_ascertaining_2018}, the merging method was applied to three canonical dynamical systems: the forced damped pendulum; the Newton fractal; and the H\'{e}non--Heiles Hamiltonian \cite{Aguirre2001}. The ``merging method'' requires as its input only  the exit basin diagrams at a finite resolution, in other words, an image of the BH shadows. The method itself is agnostic to the underlying physics or dynamics. A practical advantage of the method is that, once one has a picture of the shadows, the merging method will determine whether the shadow has the Wada property up to a certain resolution.

The paper is organized as follows. In Sec.~\ref{sec:mpbinary} we introduce the Majumdar--Papapetrou solution which describes a pair of extremally charged BHs in static equilibrium. We explore its similarities with the well-known H\'enon--Heiles Hamiltonian, and describe the exit basins in phase space. We also explain the construction of the shadows in terms of one-dimensional (1D) exit basins, and describe the role played by the so-called fundamental photon orbits. In Sec.~\ref{sec:wadamerge}, the merging method to test the Wada property is briefly reviewed. The results are presented in Sec.~\ref{sec:results}, where we apply the merging method both to the basins in phase space and to the shadows of the binary BH system. Finally, the main points of the work are summarised and discussed in Sec.~\ref{sec:discussion}.

\emph{Conventions:} The 4D spacetime metric $g_{\mu \nu}$ has Lorentzian signature $(-,+,+,+)$. The Einstein summation convention for repeated indices is assumed throughout. Indices are lowered (raised) with the metric (inverse metric), i.e., $u_{\mu} = g_{\mu \nu} u^{\nu}$ ($u^{\mu} = g^{\mu \nu} u_{\nu}$). We employ units in which the speed of light $c$ and the gravitational constant $G$ are equal to unity. Greek letters $\mu, \nu, \ldots$ denote spacetime indices from $0$ (the temporal component) to $3$; Latin letters $i, j, \ldots$ denote spatial indices from $1$ to $3$.

\section{The model: Majumdar--Papapetrou binary black hole \label{sec:mpbinary}}

\subsection{Hamiltonian formalism \label{sec:hamiltonian}}

The Majumdar--Papapetrou (MP) binary BH, or \emph{di-hole}, is a static axisymmetric solution to the Einstein--Maxwell equations of gravity and electromagnetism. The solution describes the exterior spacetime of a pair of extremally charged Reissner--Nordstr\"{o}m BHs (each with its mass parameter equal to its charge parameter: $G M/c^{2} = \sqrt{G/(4\pi \epsilon_0 c^4)} \, Q$), in static equilibrium due to the balance between their mutual gravitational attraction and electrostatic repulsion. For an overview of the MP binary BH, see e.g.~\cite{Majumdar1947,Papapetrou1945,HartleHawking1972,Chandrasekhar1989,Contopoulos1990,ShipleyDolan2016}.

The MP spacetime for a pair of equal-mass BHs is described in cylindrical coordinates $q^\mu = \{t, \rho, z, \phi\}$ by the line element
\beq \label{eqn:mpmetric}
\ed s^{2} = g_{\mu \nu} \ed q^{\mu} \ed q^{\nu} = - \frac{ \ed t^{2}}{U^{2}} + U^{2} \left( \ed \rho^{2} + \ed z^{2} + \rho^{2} \ed \phi^{2} \right),
\eeq
with electromagnetic one-form potential $A_{\mu} = \left[ 1/U, 0, 0, 0 \right]$, where
\beq
U(\rho, z) = 1 + \frac{M}{\sqrt{\rho^{2} + (z - d/2)^{2}}} + \frac{M}{\sqrt{\rho^{2} + (z + d/2)^{2}}}.
\eeq
Here, $g_{\mu \nu}$ are the covariant components of the metric tensor, $M$ is the mass of the individual BHs, and $d$ is the distance between the BHs in the background coordinates. We hereafter employ units in which $M = 1$. An artifact of the chosen coordinate system is that BH event horizons appear as single points, located on the $z$-axis at $z = \pm d/2$. These ``points'' are actually null surfaces with topology $S^{2} \times \mathbb{R}$.

The geodesics $q^{\mu} (\lambda)$ are solutions of Hamilton's equations, with Hamiltonian $H(q, p) = \frac{1}{2} g^{\mu \nu} (q) p_{\mu} p_{\nu}$, where $p_{\mu} = g_{\mu \nu} \dot{q}^{\nu}$ are the canonical momenta, $g^{\mu \nu}$ are the contravariant components of the metric tensor, and an overdot denotes differentiation with respect to the affine parameter $\lambda$.

Along geodesics, the Hamiltonian $H$ is conserved. In addition, the time-independence and axial symmetry of the Hamiltonian mean that $t$ and $\phi$ are ignorable coordinates, and $p_{t}$ and $p_{\phi}$ are constants of motion. For null geodesics (light rays), we have $H = 0$, and we may set $p_{t} = -1$ without loss of generality, as this is equivalent to rescaling the affine parameter $\lambda$.

Null geodesics are invariant under a conformal transformation of the metric of the form $g_{\mu \nu} \mapsto \Omega^{2}(q) g_{\mu \nu}$, where $\Omega(q)$ is a function of the spacetime coordinates. Performing a conformal transformation with $\Omega = U^{-1}$ allows us to express the Hamiltonian in canonical form as
\begin{eqnarray}
H &=& \frac{1}{2} (p_{\rho}^{2} + p_{z}^{2}) + V = 0, \\
V (\rho, z) &=& -\frac{1}{2\rho^{2}} ( h - p_{\phi} ) ( h + p_{\phi} ),
\end{eqnarray}
where, in order to factorize the potential $V(\rho, z)$, we have introduced the \emph{height function} (or \emph{effective potential})
\beq \label{eqn:height_function}
h(\rho, z) = \rho U^{2}.
\eeq
The so-called \emph{null condition} $H = 0$, and the positivity of the kinetic term in the Hamiltonian together imply that $V \leq 0$. This inequality defines the \emph{allowed regions} in configuration space; the solutions of $h = \pm p_{\phi}$ (which are equivalent to $V = 0$) define the boundary of the allowed regions.

The full phase space is 8D, spanned by the four spacetime coordinates and their conjugate momenta $\{q^{\mu}, p_{\mu} \}$. However, the conserved quantities allow us to focus on a reduced 4D phase space with two pairs of conjugate variables, $\{ \rho, z, p_{\rho}, p_z \}$, and one constraint $H=0$. The null constraint $H = 0$ allows us to express one coordinate, e.g.~$p_{z}$, in terms of the other three coordinates.

\begin{figure}
\subfigure[H\'{e}non--Heiles]{\includegraphics[height=7cm]{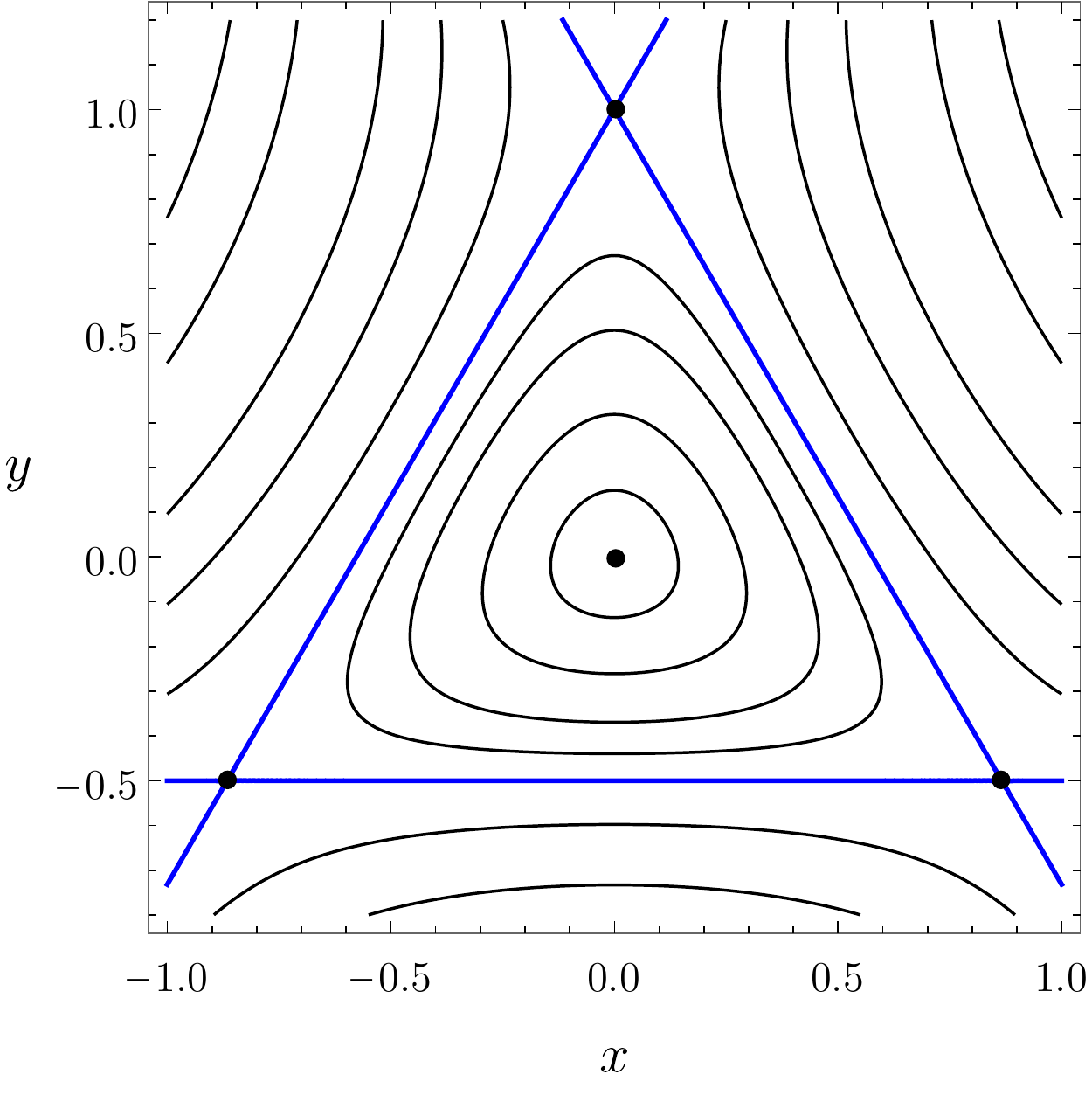} \label{fig:hhequipotential}} \hspace{1cm}
\subfigure[Majumdar--Papapetrou]{\includegraphics[height=7cm]{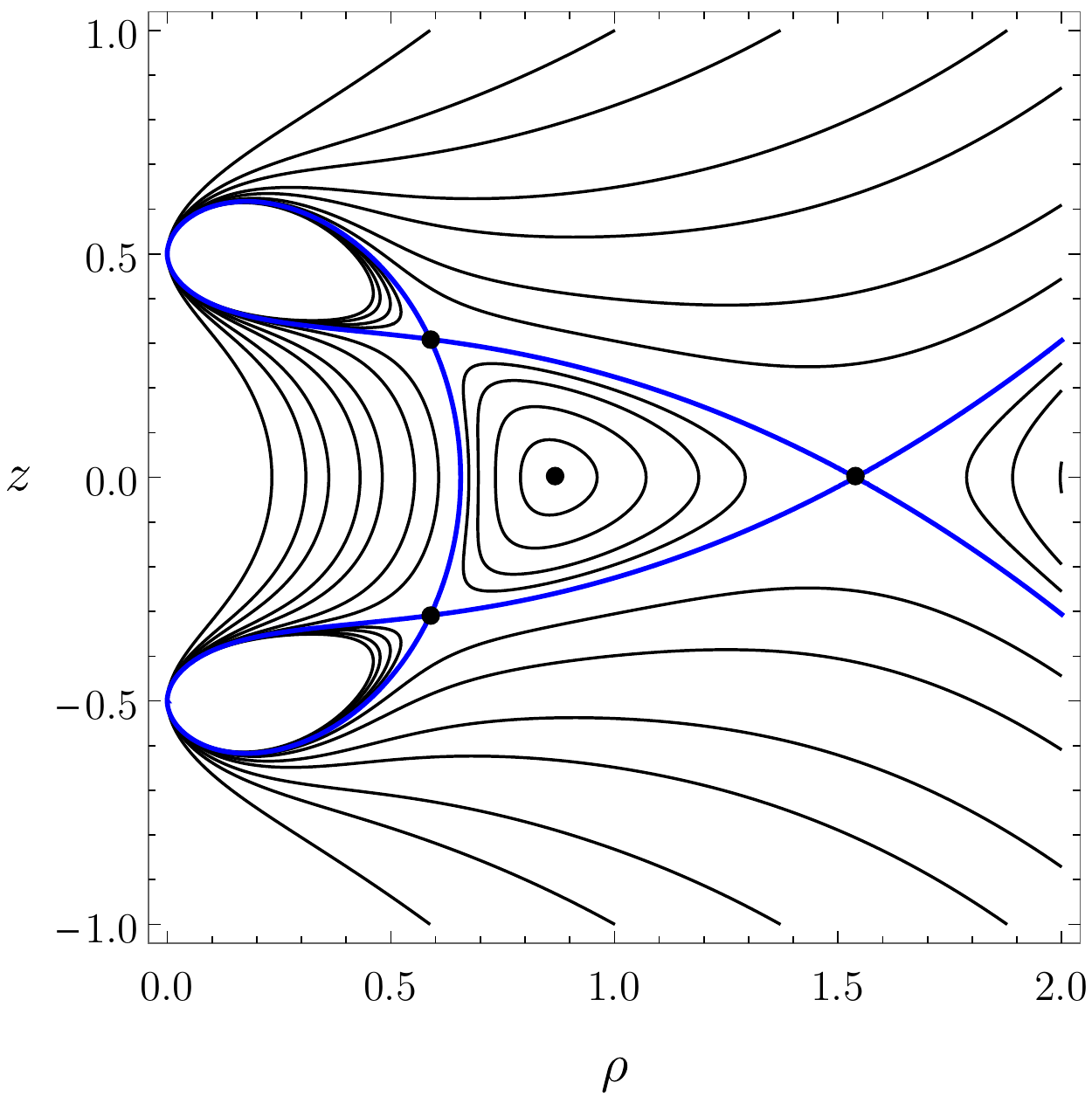} \label{fig:mpequipotential}}
\caption{Equipotential curves for (a) the HH potential in the $(x,y)$-plane; and (b) the MP height function (effective potential) in the $(\rho,z)$-plane for $d = M = 1$. In both cases, the critical contour which connects three saddle points is shown in blue. The critical contour encloses (a) a minimum at $(0,0)$, and (b) a maximum at $(\sqrt{3}/2,0)$. Both the HH and MP systems can either be open with three escapes, or closed. See the text for details.}
 \label{fig:equipotentials}
\end{figure}

The MP di-hole system, in the reduced phase space, has features in common \cite{ShipleyDolan2016,DolanShipley2016} with the H\'{e}non--Heiles (HH) Hamiltonian system \cite{HenonHeiles1964}, which has become a paradigm for 2D time-independent Hamiltonian scattering. The HH model, first introduced to study galactic dynamics, has the Hamiltonian
\begin{eqnarray}
H_{\mathrm{HH}} &=& \frac{1}{2} \left( p_{x}^{2} + p_{y}^2 \right) + V_{\mathrm{HH}} = E, \\ V_{\mathrm{HH}}(x, y) &=& \frac{1}{2} ( x^{2} + y^{2} ) + x^{2} y - \frac{1}{3}y^{3},
\end{eqnarray}
where $E$ is the total energy of the system. The MP di-hole system in its reduced phase space and the HH system above are both examples of 2D time-independent Hamiltonian systems with a single free parameter for rays: $p_{\phi}$ in the former case, and $E$ in the latter. Figure \ref{fig:equipotentials} shows equipotential curves for (a) the HH Hamiltonian and (b) the equal-mass MP dynamical system with $d = 1$. In both cases, there is a ``critical contour'' connecting three saddle points, which encloses an unstable fixed point.

The HH Hamiltonian is invariant under $2\pi/3$ rotations. It has four fixed points, where $\bnab V_{\mathrm{HH}} = (0,0)$: a minimum at $(x,y) = (0,0)$, and three saddle points at $(x,y) = (0,1)$ and $(x,y) = (\pm \sqrt{3}/2, -1/2)$. The three saddle points are connected by a single equipotential curve, with critical energy $E^{\ast} = 1/6$. The contour $V_{\mathrm{HH}} = E^{\ast}$ encloses the minimum, at which $E = 0$. For energies below the critical value ($E \leq E^{\ast}$), the HH system is \emph{closed}. However, for energies above the critical value ($E > E^{\ast}$), the HH system is \emph{open}, with three escapes connecting the scattering region to infinity.
The HH system is investigated in the context of chaotic scattering, where orbits can escape from the scattering region, in \cite{Aguirre2001}. For a comprehensive review of fractal structures in the exit basins of open Hamiltonian systems, see \cite{Aguirre2009}.

For the particular case of equal-mass BHs separated by coordinate distance $d = 1$, the MP di-hole shares key qualitative features with the HH system \cite{ShipleyDolan2016,DolanShipley2016} (see Fig.~\ref{fig:equipotentials}). There are three saddle points, one of which is in the equatorial plane ($z = 0$) at $(\rho, z) = (\frac{1}{2} 5^{1/4} \varphi^{3/2}, 0)$, and the other two are out of the plane, at $(\rho, z) = (\frac{1}{2} 5^{1/4} \varphi^{-1/2}, \pm (2 \varphi)^{-1})$, where $\varphi = \frac{1}{2}(1 + \sqrt{5})$ denotes the golden ratio. The three saddle points are connected by a single critical contour, $h = p_{\phi}^{\ast} = \frac{1}{2} 5^{5/4} \varphi^{3/2}$. The critical contour encloses a local maximum of $h$ at $(\rho, z) = (\sqrt{3}/2, 0)$ with $p_{\phi} = 9 \sqrt{3}/2$.

For $p_\phi$ above the threshold value ($p_{\phi} \geq p_{\phi}^{\ast}$), equipotential lines form closed curves on a subregion of $(\rho,z)$-space, and a second disconnected component of the contour extends to spatial infinity. Thus, there are two disconnected regions in the phase space. Conversely, for $p_\phi$ below the threshold ($p_{\phi} < p_{\phi}^{\ast}$), the MP system is \emph{open}, with two escapes leading to each of the BHs, and the other connecting the scattering region to spatial infinity. (Note here that, due to the fact $h$ possesses a maximum rather than a minimum, the inequalities describing the open/closed system are reversed when compared with those of the HH system.)

In Ref.~\cite{DolanShipley2016}, the authors elucidate the similarities between the \emph{closed} HH and MP systems, analysing the transition from regularity to chaos through the use of Poincar\'{e} sections. In this article, we discuss the MP di-hole as a novel example of a 2D time-independent Hamiltonian system with three escapes.

\subsection{Exit basins in phase space \label{sec:exitbasins}}

In open Hamiltonian systems with multiple escapes, one can define \emph{exit basins} in a similar way to the basins of attraction in a dissipative system. An exit basin is defined as the set of initial conditions which lead to a certain escape in the future. This is realised by numerically integrating the equations of motion for a fine grid of initial conditions. Each trajectory is integrated until it leaves the scattering region through one of the exits. The initial conditions are then divided into basins according to the asymptotic state.

The exit basins for the HH system were introduced and studied extensively in Ref.~\cite{Aguirre2001}. The authors applied computational methods to verify that the basins of the HH system possess the Wada property; that is, each point on a basin boundary is on the boundary of all three basins. Below we shall consider rays in the MP spacetime whose initial conditions are defined in close analogue with the HH study \cite{Aguirre2001}.

\begin{figure}
\subfigure[$(\rho, z)$-space]{\includegraphics[height=7cm]{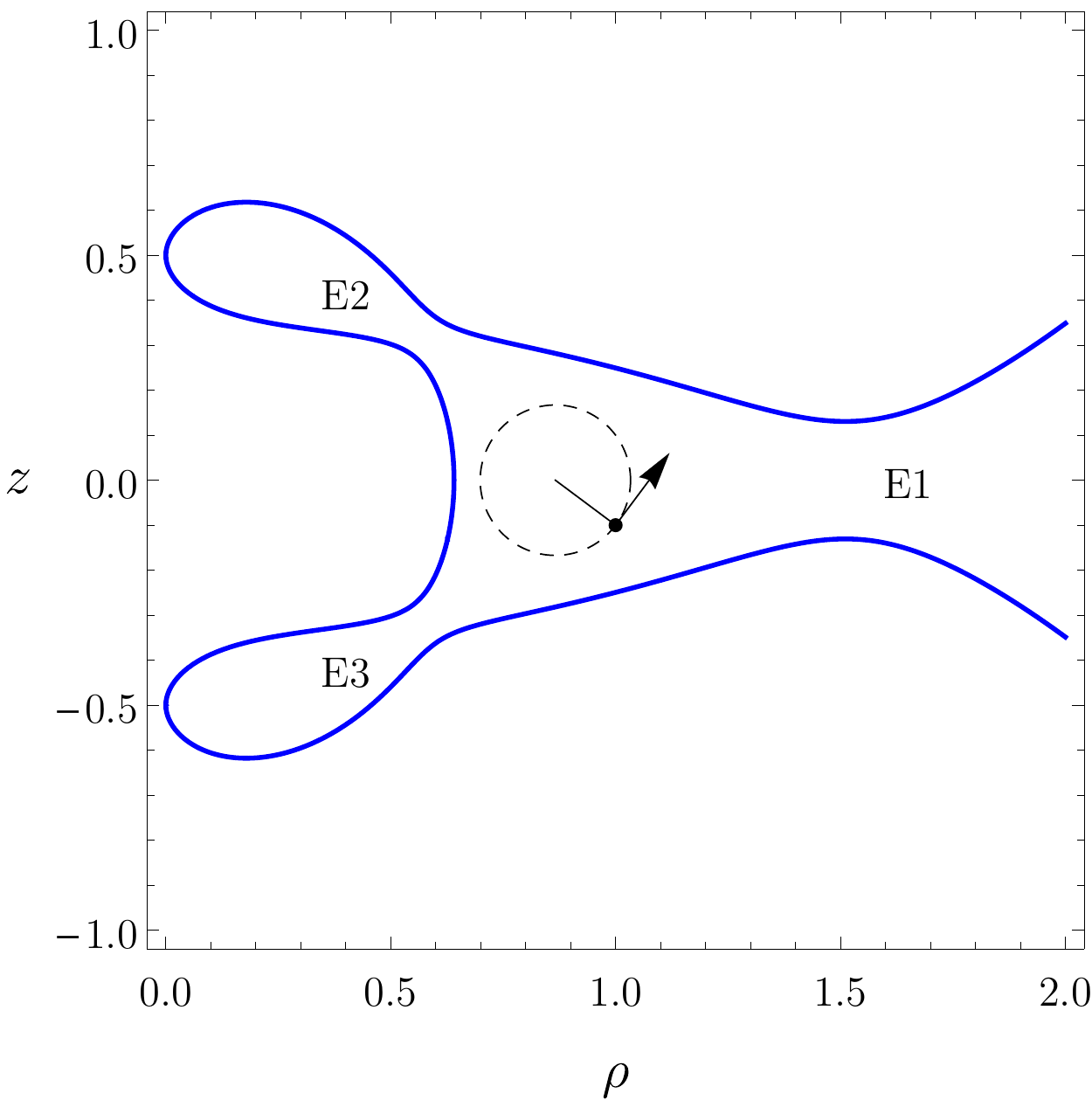} \label{fig:mp_initial_conditions_rho_z}} \hspace{1cm}
\subfigure[$(\rho, p_{\rho})$-space]{\includegraphics[height=7cm]{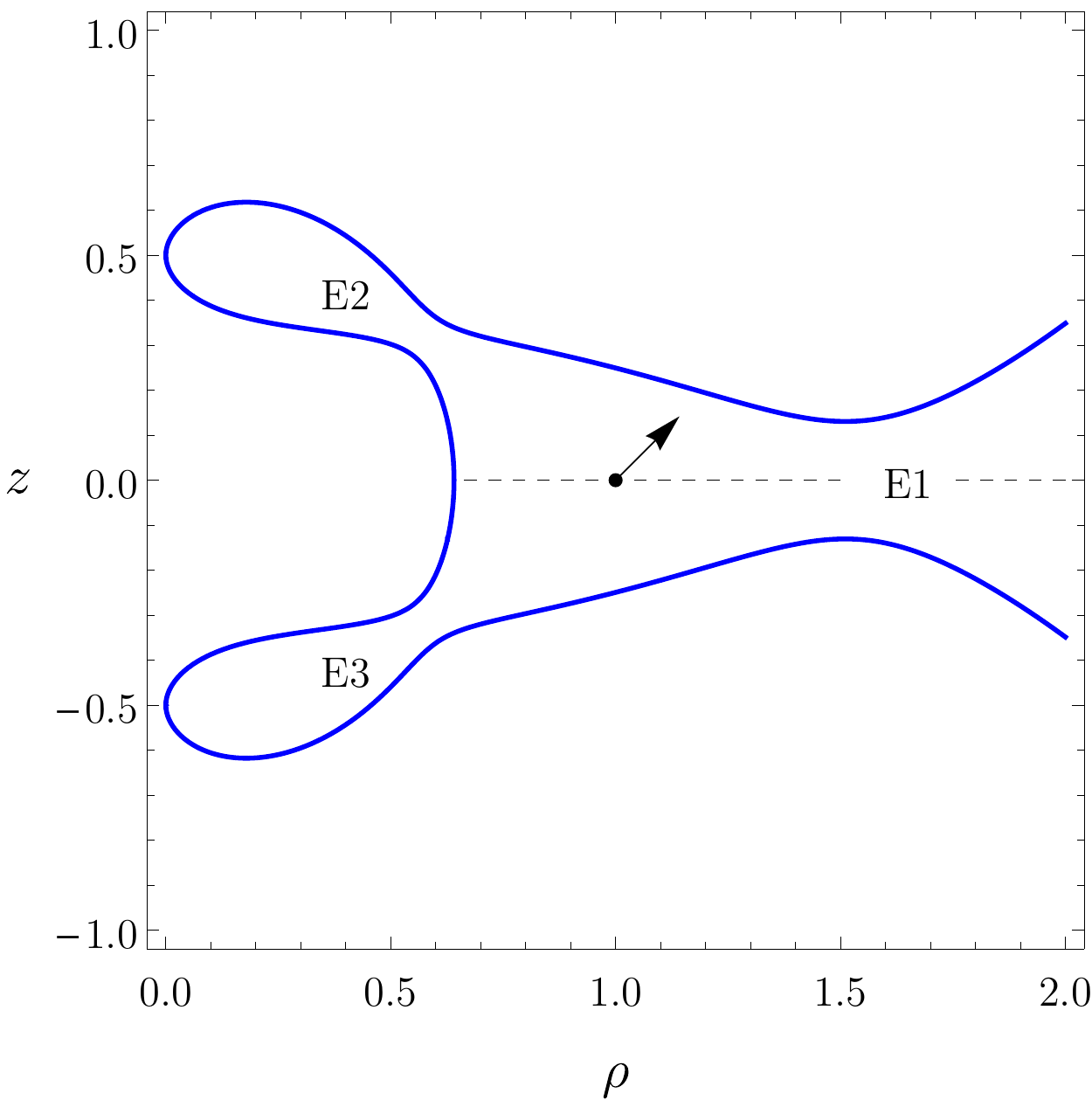} \label{fig:mp_initial_conditions_rho_prho}}
\caption{Choice of initial conditions used to plot the exit basin diagrams. (a) Initial conditions in $(\rho, z)$-space. The photon has initial three-momentum tangent to the circle centred on the maximum of $h$. (b) Initial conditions in $(\rho, p_{\rho})$-space. The photon is fired from the $\rho$-axis ($z = 0$) and the value of $p_{\rho}$ is varied. In both cases, the blue curve is the contour $h = p_{\phi}^{\ast} - \Delta p_{\phi}$, with $\Delta p_{\phi} = 0.02$. \label{fig:mp_initial_conditions}}
\end{figure}

\begin{figure*}
\centering
\subfigure[$\Delta p_{\phi} = 0.01$]{\includegraphics[width=0.32\textwidth]{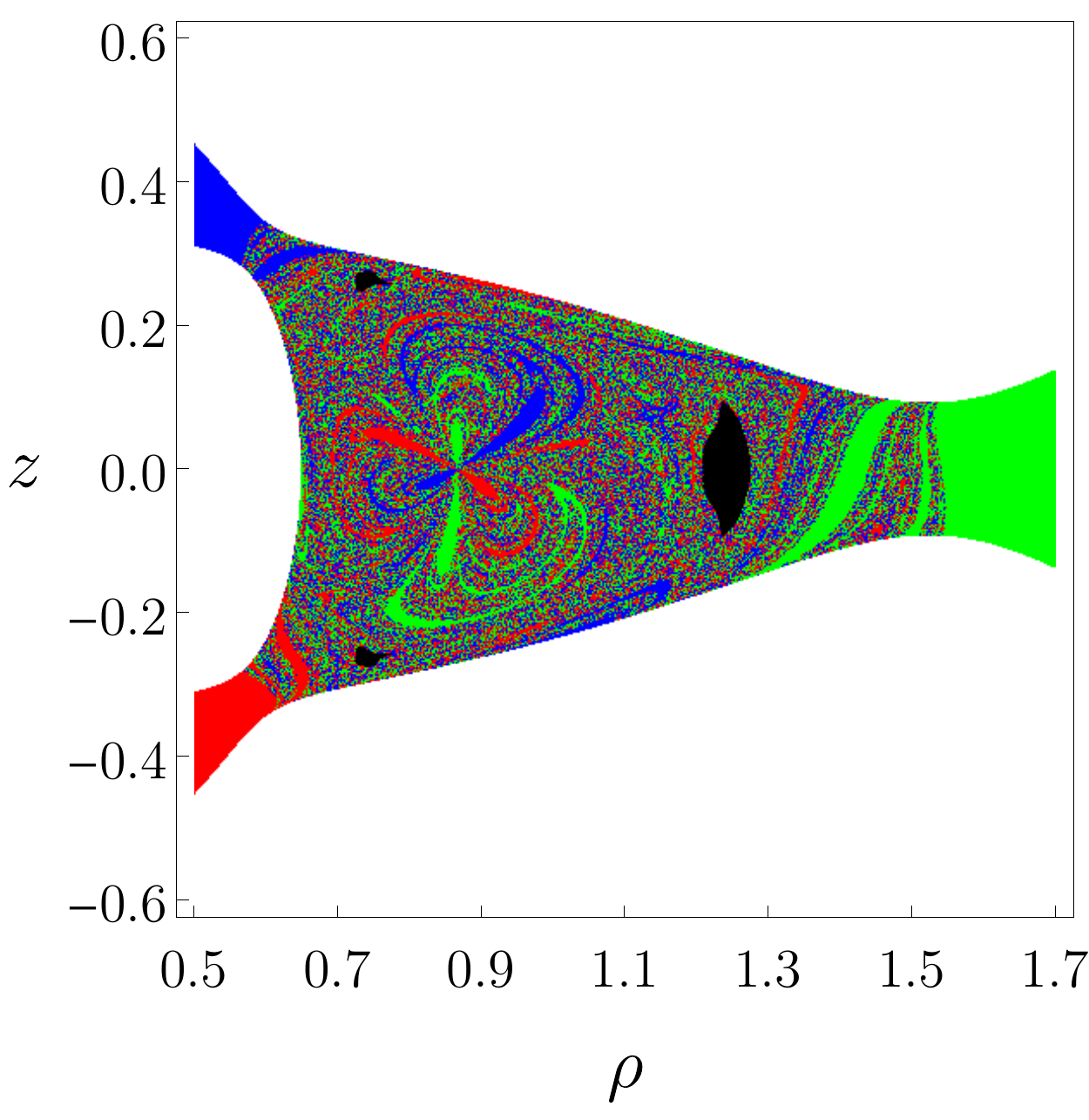} \label{fig:mp_basins_rho_z_1}}
\subfigure[$\Delta p_{\phi} = 0.03$]{\includegraphics[width=0.32\textwidth]{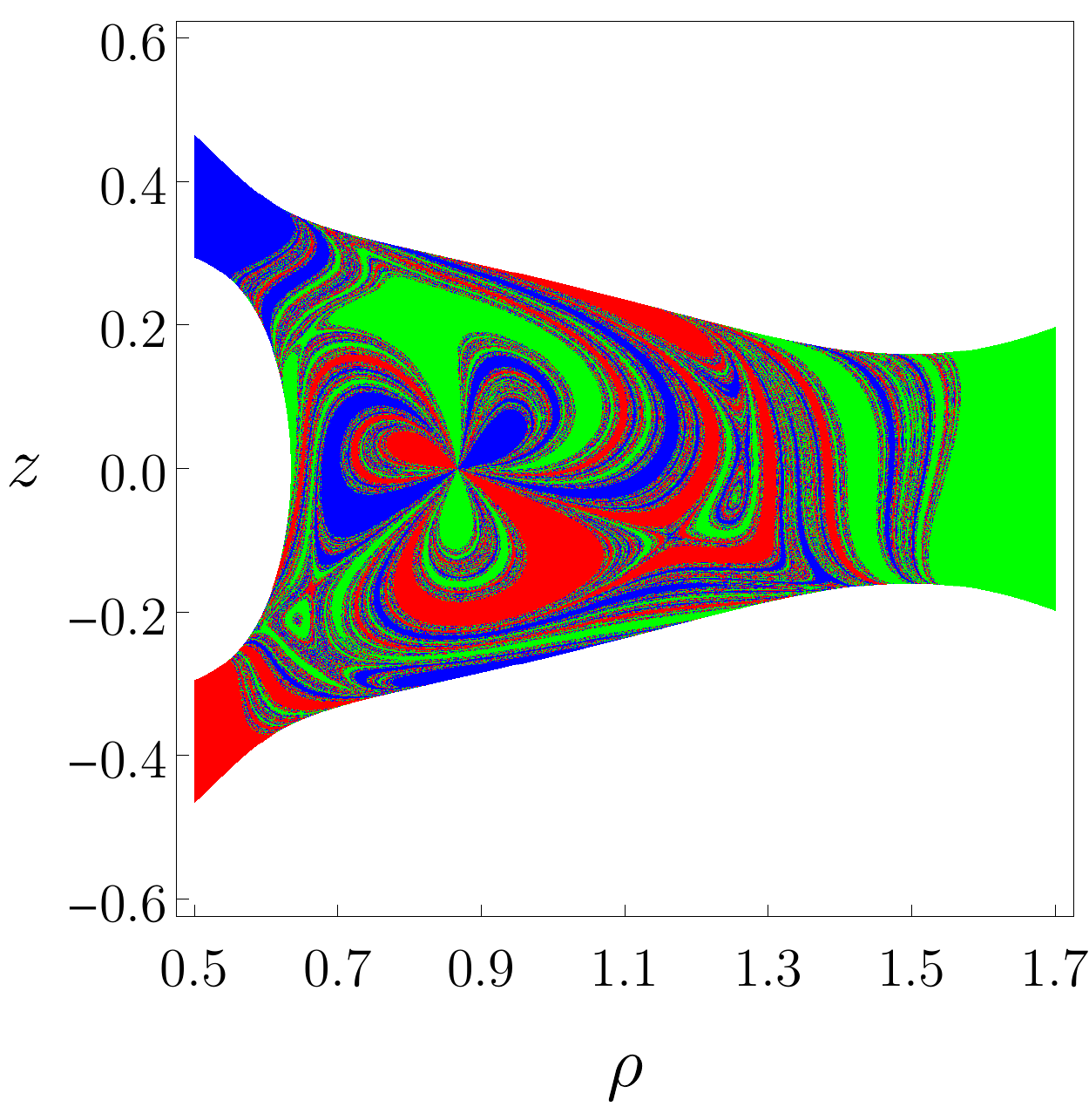} \label{fig:mp_basins_rho_z_2}}
\subfigure[$\Delta p_{\phi} = 0.05$]{\includegraphics[width=0.32\textwidth]{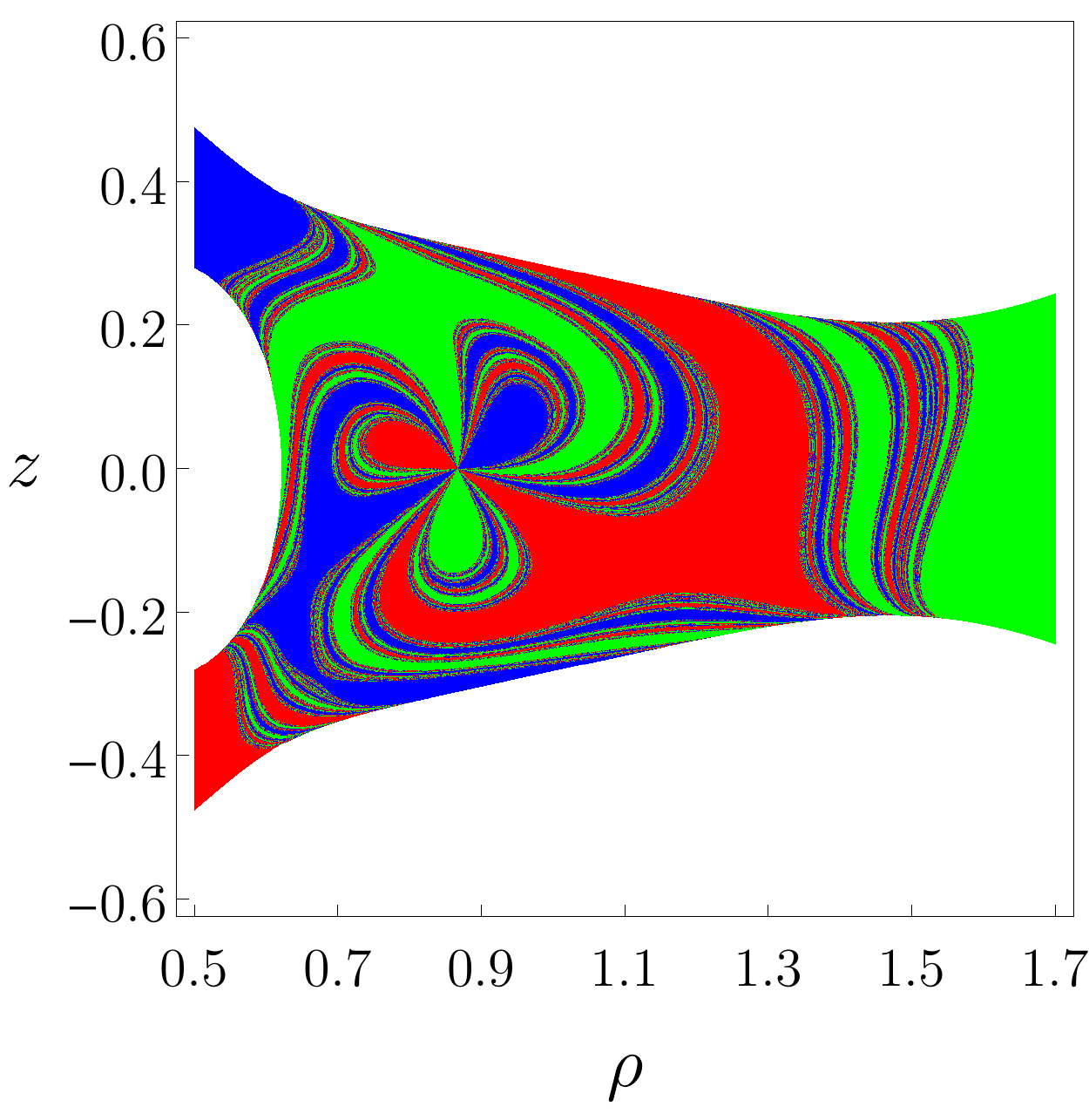} \label{fig:mp_basins_rho_z_3}}
\subfigure[$\Delta p_{\phi} = 0.01$]{\includegraphics[width=0.32\textwidth]{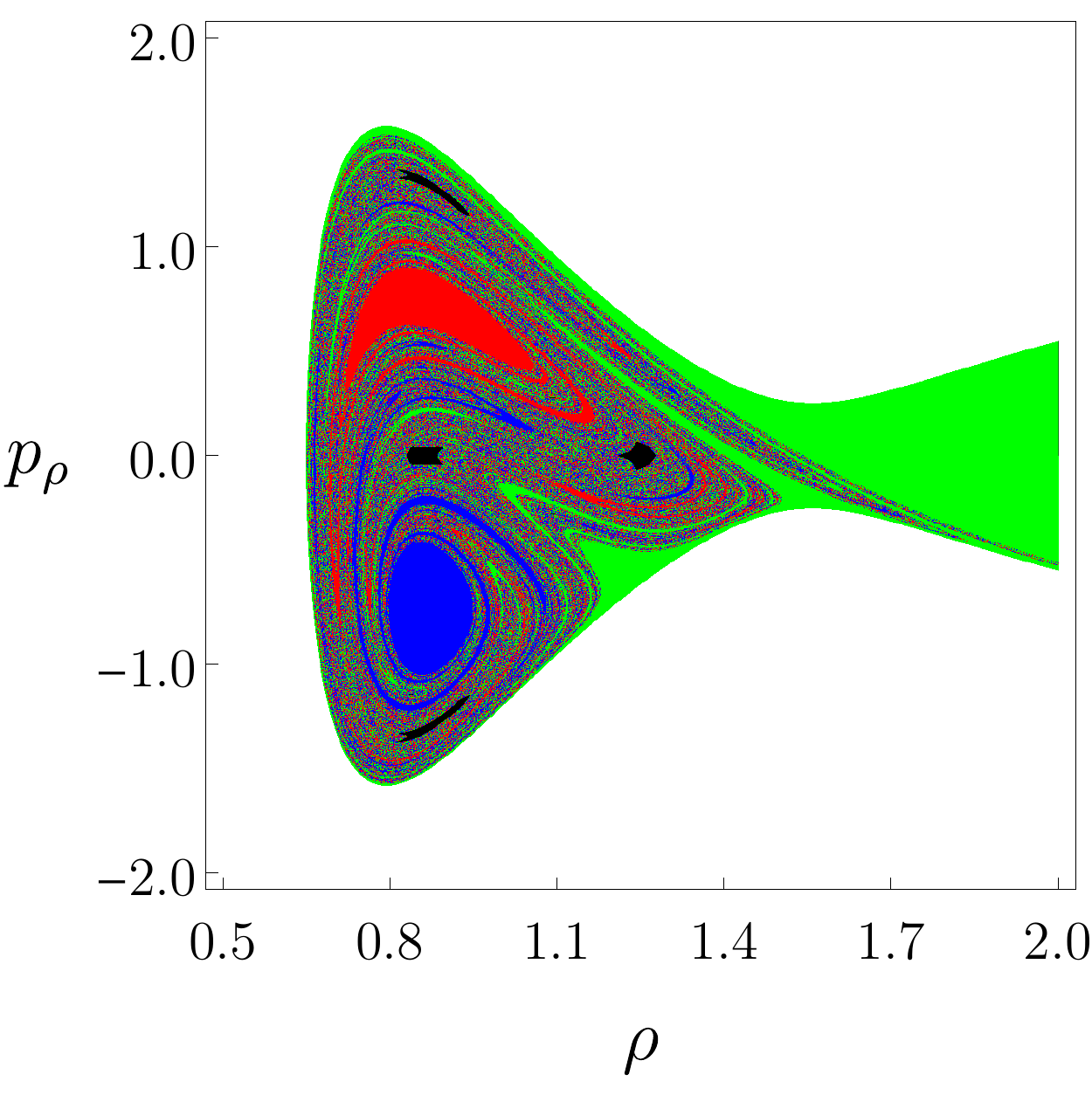} \label{fig:mp_basins_rho_prho_1}}
\subfigure[$\Delta p_{\phi} = 0.03$]{\includegraphics[width=0.32\textwidth]{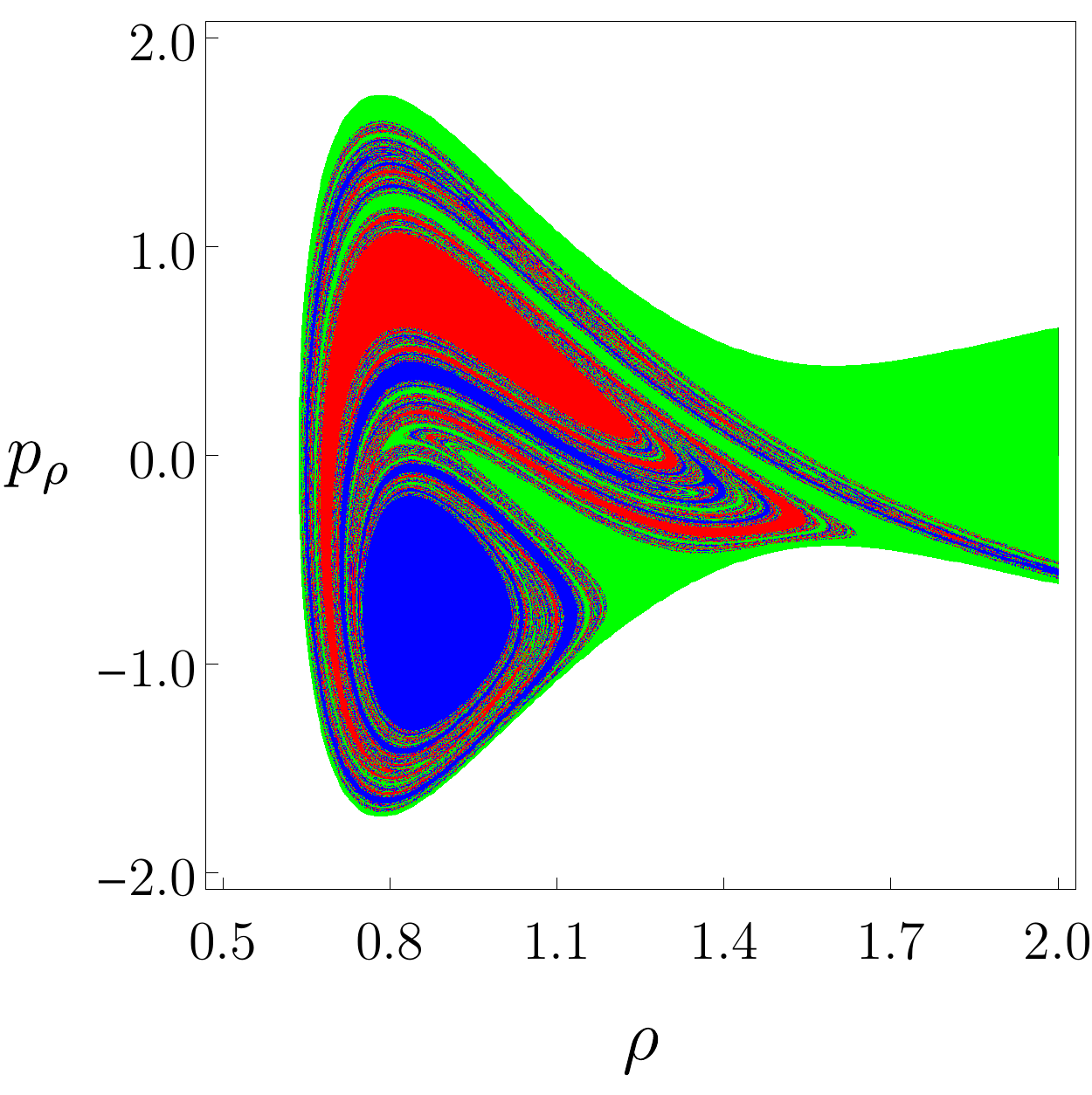} \label{fig:mp_basins_rho_prho_2}}
\subfigure[$\Delta p_{\phi} = 0.05$]{\includegraphics[width=0.32\textwidth]{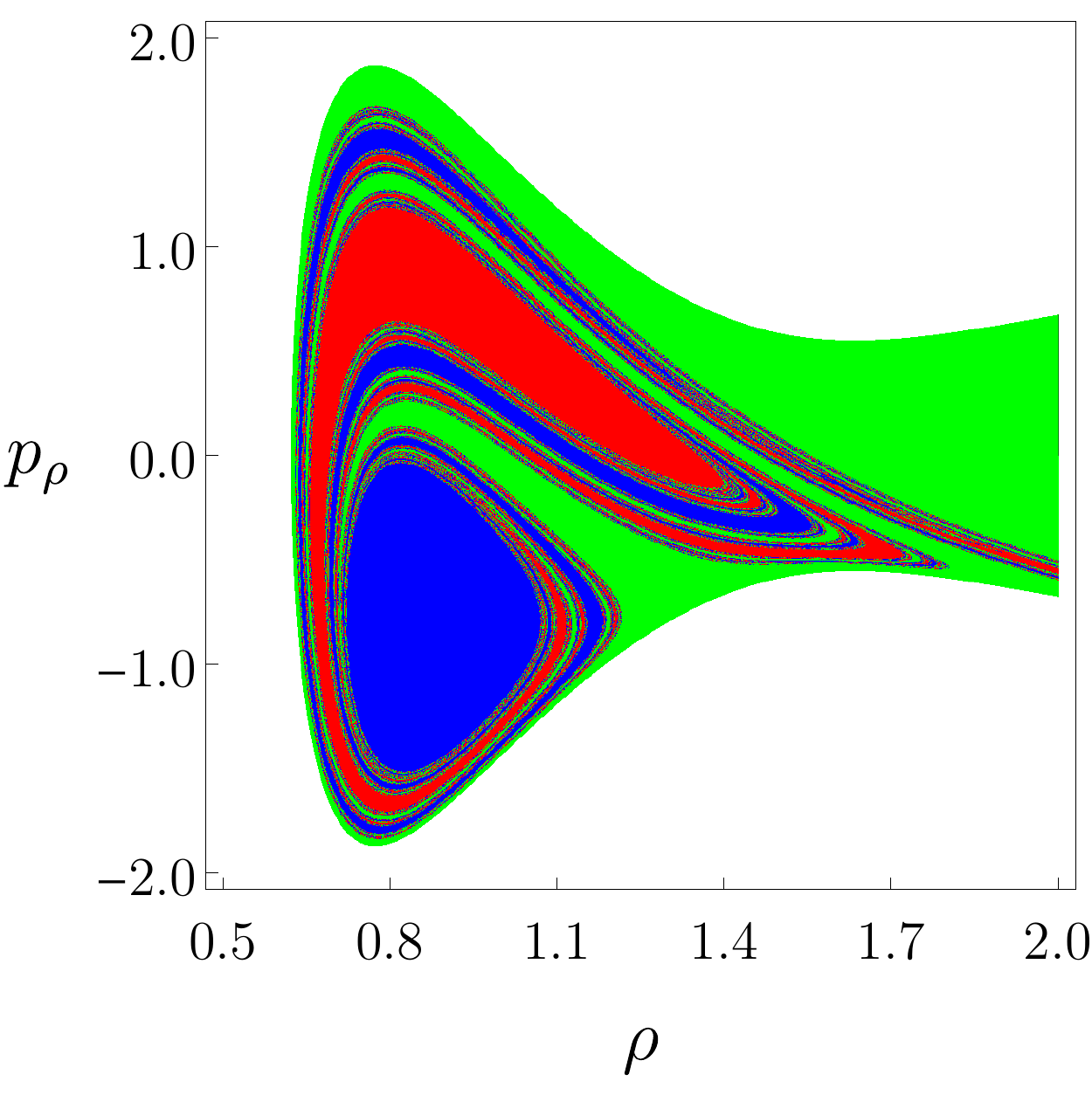} \label{fig:mp_basins_rho_prho_3}}
\caption{Exit basins for the MP di-hole for (a)--(c) the initial conditions in $(\rho, z)$-space, presented in Fig.~\ref{fig:mp_initial_conditions_rho_z}; and (d)--(f) the initial conditions in $(\rho, p_{\rho})$-space, presented in Fig.~\ref{fig:mp_initial_conditions_rho_prho}. In each case, the initial data which lead to infinity are plotted in green, and those which asymptote to the event horizon of the upper (lower) BH are shown in blue (red). As $\Delta p_{\phi} \equiv p_{\phi}^{\ast} - p_{\phi}$ decreases, the escape width decreases and KAM islands of stability (plotted in black) dominate the phase space (see text). \label{fig:mp_basins}}
\end{figure*}

The first choice of initial conditions is to fix the coordinates $\rho$ and $z$, and choose the initial three-momentum to be tangential (in the anticlockwise sense) to the circle of radius $\sqrt{(\rho - \rho_{\mathrm{max}})^{2} + (z - z_{\mathrm{max}})^{2}}$, centred on the maximum of $h$, which is located at $(\rho_{\mathrm{max}}, z_{\mathrm{max}}) = (\sqrt{3}/2,0)$. The exit basins are then plotted in the $(\rho, z)$-plane. See Fig.~\ref{fig:mp_initial_conditions_rho_z} for the set-up and Figs.~\ref{fig:mp_basins_rho_z_1}--\ref{fig:mp_basins_rho_z_3} for the corresponding exit basin diagrams.

Our second choice of the initial conditions is to fix $z=0$, and then vary the values of $\rho$ and $p_{\rho}$. The exit basin diagrams are plotted in the $(\rho, p_{\rho})$-plane. The initial conditions are shown in Fig.~\ref{fig:mp_initial_conditions_rho_prho}, and the exit basins are shown in Figs.~\ref{fig:mp_basins_rho_prho_1}--\ref{fig:mp_basins_rho_prho_3}.

In order to visualise the exit basins, we colour the initial conditions green if they lead to the attractor at infinity, blue for the upper BH, and red for the lower BH. The Kolmogorov--Arnold--Moser (KAM) islands of stability \cite{Tabor1989,Contopoulos2002} are plotted in black. These KAM tori comprise the set of initial conditions corresponding to orbits which never escape the scattering region as $\lambda \rightarrow \pm \infty$, despite the fact that $p_{\phi} < p_{\phi}^{\ast}$, i.e., the system is open. Trajectories inside the KAM islands never escape to infinity nor end up in either of the BHs; rather, they keep wandering forever with a quasiperiodic motion. The KAM islands of stability are organized in a fractal hierarchy and they have a non-zero measure, as can be inferred from the black regions depicted in Figs.~\ref{fig:mp_basins_rho_z_1} and \ref{fig:mp_basins_rho_prho_1}. As one increases $\Delta p_{\phi}$, all trajectories escape the scattering region and the KAM islands disappear. A similar effect occurs in the H\'{e}non--Heiles Hamiltonian as the energy is increased \cite{BarrioBlesaSerrano2008}. Examples of quasiperiodic non-escaping orbits for the equal-mass MP di-hole are shown in Fig.~4 of Ref.~\cite{DolanShipley2016}. For a general discussion of the limit of small escapes in open Hamiltonian systems we refer the reader to \cite{AguirreSanjuan2003}.

Increasing the value of $\Delta p_{\phi} \equiv p_{\phi}^{\ast} - p_{\phi}$ increases the width of the three escapes. In Fig.~\ref{fig:mp_basins}, we plot the MP basins for a selection of values of $\Delta p_{\phi}$. As the value of $\Delta p_{\phi}$ increases, the KAM islands disappear, and the basins, both in the $(\rho, z)$ and the $(\rho, p_{\rho})$ subspaces become visibly less fractalised (this effect could be quantified by using the fractal dimension or the basin entropy \cite{DazaWagemakersGeorgeotEtAl2016}). By inspection of the basins, one can see striking similarities between the MP di-hole system and those of the HH Hamiltonian, which are presented in e.g.~Figs.~4 and 5 of \cite{Aguirre2001}.

\subsection{Black hole shadows \label{sec:shadows}}

\begin{figure}[h]
\includegraphics[height=7cm]{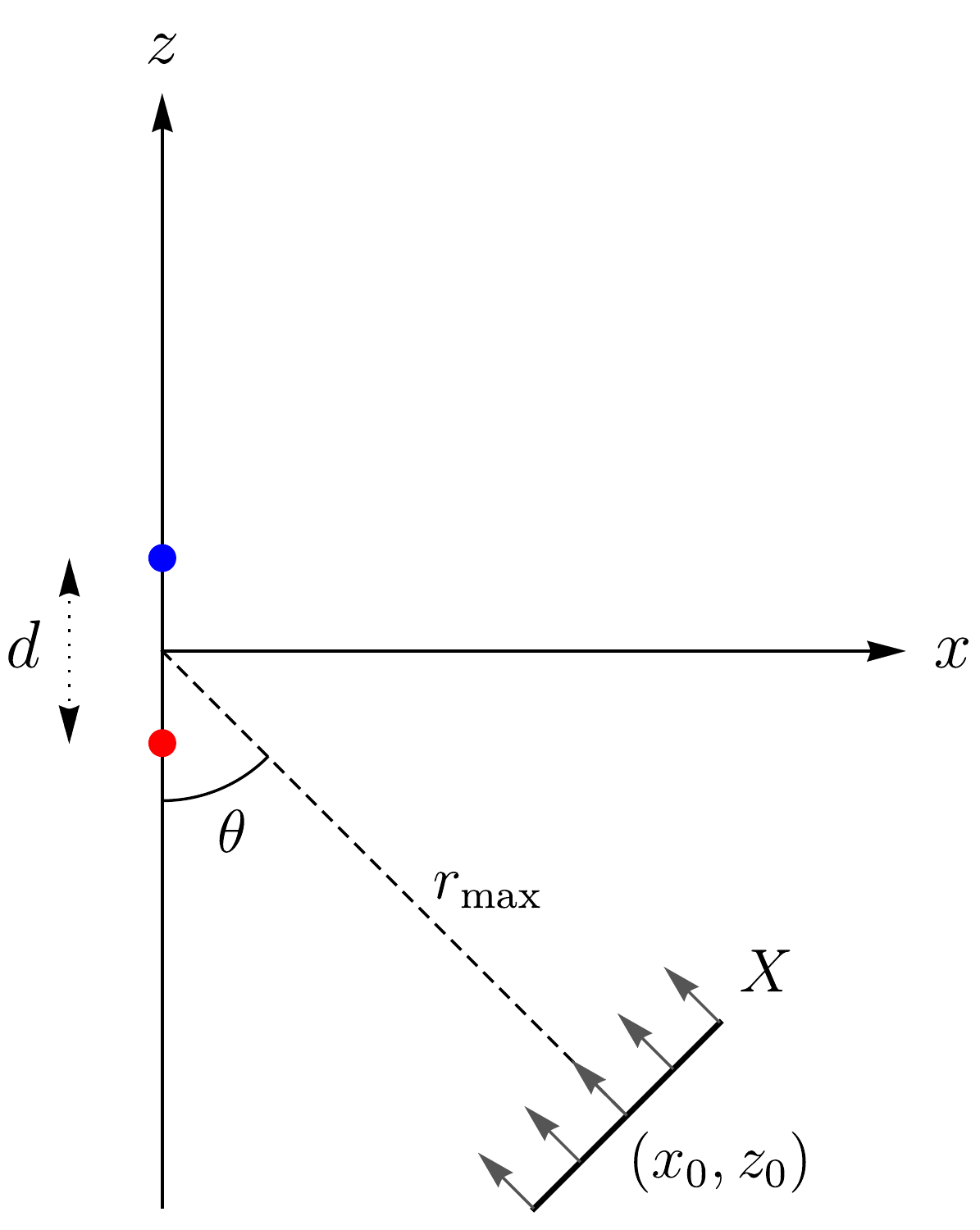}
\caption{Schematic diagram of the ray-tracing algorithm used to compute MP di-hole shadows in the $(x,z)$-plane. The BHs are located at $z = \pm d/2$, separated by coordinate distance $d$. The upper (lower) BH is represented using a blue (red) circle. The observer's image plane is located at $r = \sqrt{\rho^{2} + z^{2}} = r_{\textrm{max}}$ with viewing angle $\theta$, and is spanned by the image plane coordinates $(X,Y)$; the $Y$-direction is suppressed in the diagram. The relationship between the background coordinates and the observer's coordinates is explained in the text.} \label{fig:mp_shadow_set_up}
\end{figure}

A BH shadow is defined with respect to a family of rays on an initial data surface. Here we consider rays which pass orthogonally through a planar surface with centre $(\rho_{0},z_{0})$, where $\sqrt{\rho_{0}^{2} + z_{0}^{2}} = r_{\mathrm{max}}$. We typically take $r_{\mathrm{max}} = 50$, which is sufficiently far from the system to represent the perspective of a distant observer. The observer's ``viewing angle'' $\theta$, is defined via $\sin \theta = \rho_{0}/r_{\mathrm{max}}$, $\cos \theta = -z_{0}/r_{\mathrm{max}}$. A schematic diagram of this set-up is shown in Fig.~\ref{fig:mp_shadow_set_up}.

A point (or ``pixel'') on the image plane has coordinates $(X, Y)$, related to the cylindrical coordinates via $\rho^{2} = (\rho_{0} + X \cos \theta)^{2} + Y^{2}$, $z = z_{0} - X \sin \theta$. The background Cartesian coordinates $(x, y, z)$, with $\rho^{2} = x^{2} + y^{2}$, and the image plane coordinates $(X, Y)$ are related via $x = x_{0} + X \cos \theta$, $y = Y$, $z = z_{0} - X \sin \theta$, where $(x_{0}, 0, z_{0})$ is the location of the centre of the image plane in Cartesian coordinates. There is a one-to-one correspondence between a pixel on the image plane and a null geodesic. The pixel is part of the BH shadow if and only if the corresponding geodesic approaches the event horizon of a BH when traced backwards in time.

In our setup, the image plane defines a set of nearby observers at each point $(X, Y)$. One can instead define a BH shadow with respect to a single observer, by tracing rays from a single point in spacetime, by varying the elevation and azimuth. The two definitions are essentially equivalent in the limit $r_{\text{max}} \rightarrow \infty$.

\begin{figure*}
\subfigure[$d = 1$]{\includegraphics[height=7cm]{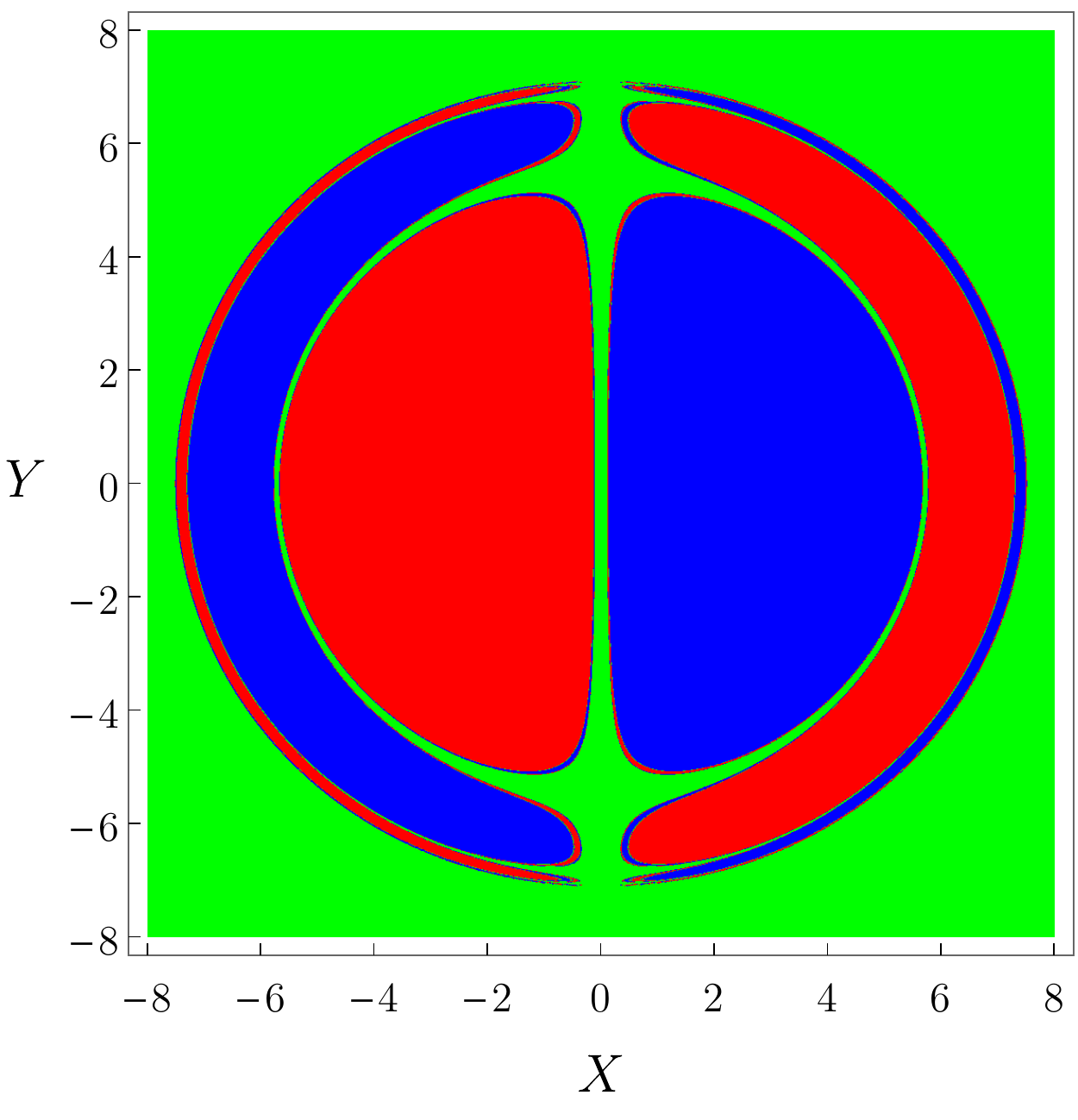}} \hspace{1cm}
\subfigure[$d = 2$]{\includegraphics[height=7cm]{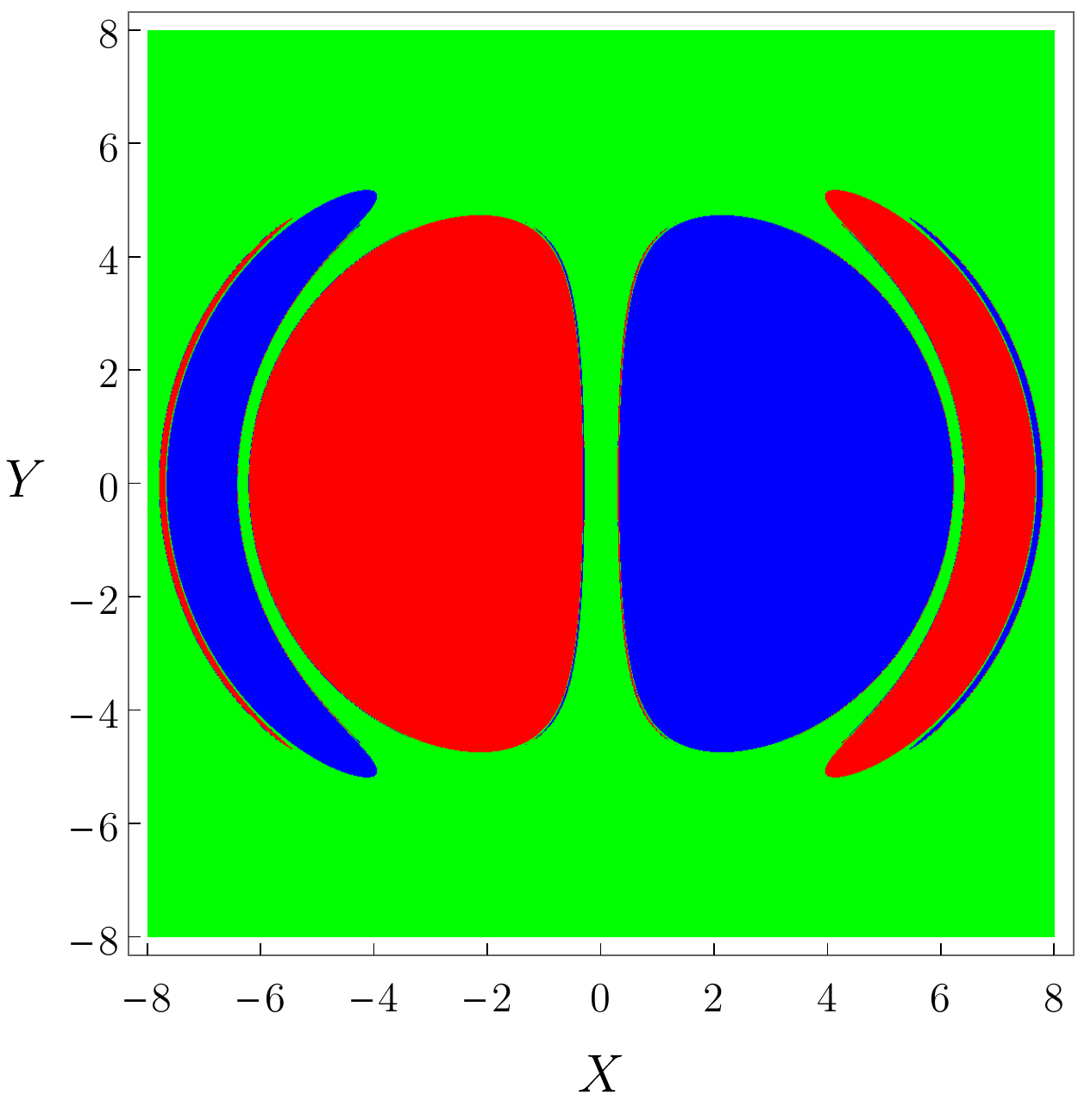}}
\caption{Shadows cast by the static MP binary BH for different values of the separation $d$. The photons which escape to spatial infinity are plotted in green; the shadow cast by the upper (lower) BH is plotted in blue (red). These three open sets can be viewed as exit basins, defined on the image plane of a distant observer.} \label{fig:mpshadows}
\end{figure*}

Figure \ref{fig:mpshadows} shows MP di-hole shadows for separations $d = 1$ and $d = 2$ and viewing angles $\theta = \pi/2$ (see Ref.~\cite{ShipleyDolan2016} for a gallery of MP shadows with a selection of viewing angles and separations). The initial conditions on the $(X, Y)$-plane which lead to the upper (lower) BH are coloured blue (red), and those which escape to infinity are coloured green.

The binary BH image (or exit basin diagram) features a pair of globular shadows corresponding to the individual BHs. Around these primary shadows, there is a self-similar hierarchy of eyebrow-like features. The boundary of the MP binary BH shadow corresponds to the set of initial conditions which asymptote towards unstable perpetual orbits. In Ref.~\cite{ShipleyDolan2016}, it was shown that these perpetual orbits form a Cantor-like set.

The 2D binary BH shadow can be viewed as a set of 1D binary BH shadows, each of which corresponds to a fixed value of the parameter $p_{\phi}$. Under a change of coordinates $x^{\mu} \mapsto x^{\mu^{\prime}}$, the momenta $p_{\mu}$ transform according to $p_{\mu} \mapsto p_{\mu^{\prime}} = \frac{\partial x^{\mu}}{\partial x^{\mu^{\prime}}} p_{\mu}$. The standard definition of cylindrical polar coordinates therefore gives the relationship $p_{\phi} = x p_{y} - y p_{x}$. A photon with initial three-momentum orthogonal to the image plane has $p_{x} = 0$ and $p_{y} = - U^{2}$; hence, $p_{\phi} = Y U^{2}$. Moreover, in the far-field limit ($r_{\mathrm{max}} \rightarrow \infty$), we have $U \rightarrow 1$. A scattering problem with $p_{\phi} = \textrm{constant}$ therefore admits a 1D shadow with $Y = \textrm{constant}$ (i.e., a horizontal slice across a 2D shadow image).

\begin{figure*}
\begin{tabular}{cccc}
\subfigure[$p_{\phi}=4$]{\includegraphics[width=3.8cm]{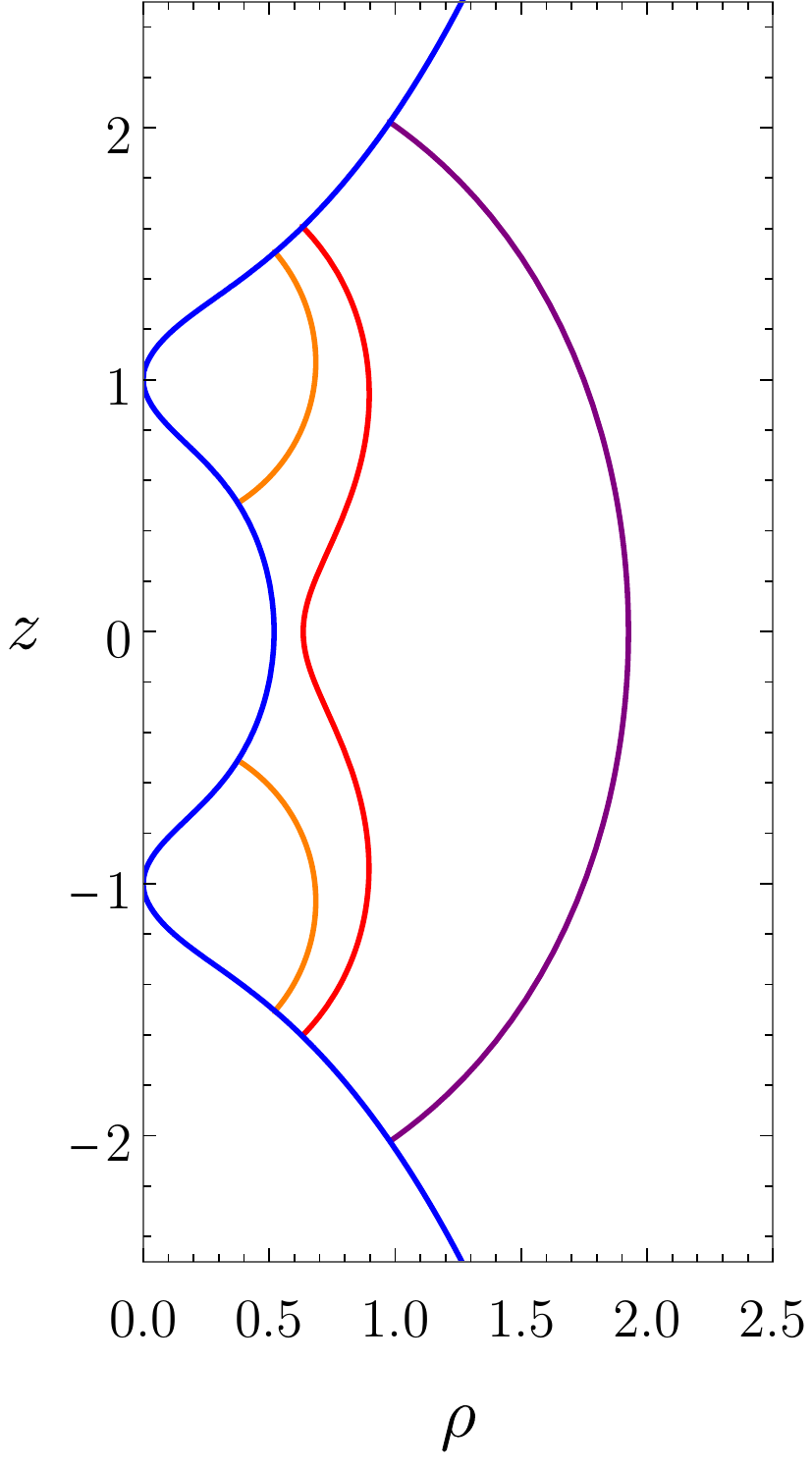} \label{fig:mp_fundamental_orbits_a}}
&
\subfigure[$p_{\phi}=5$]{\includegraphics[width=3.8cm]{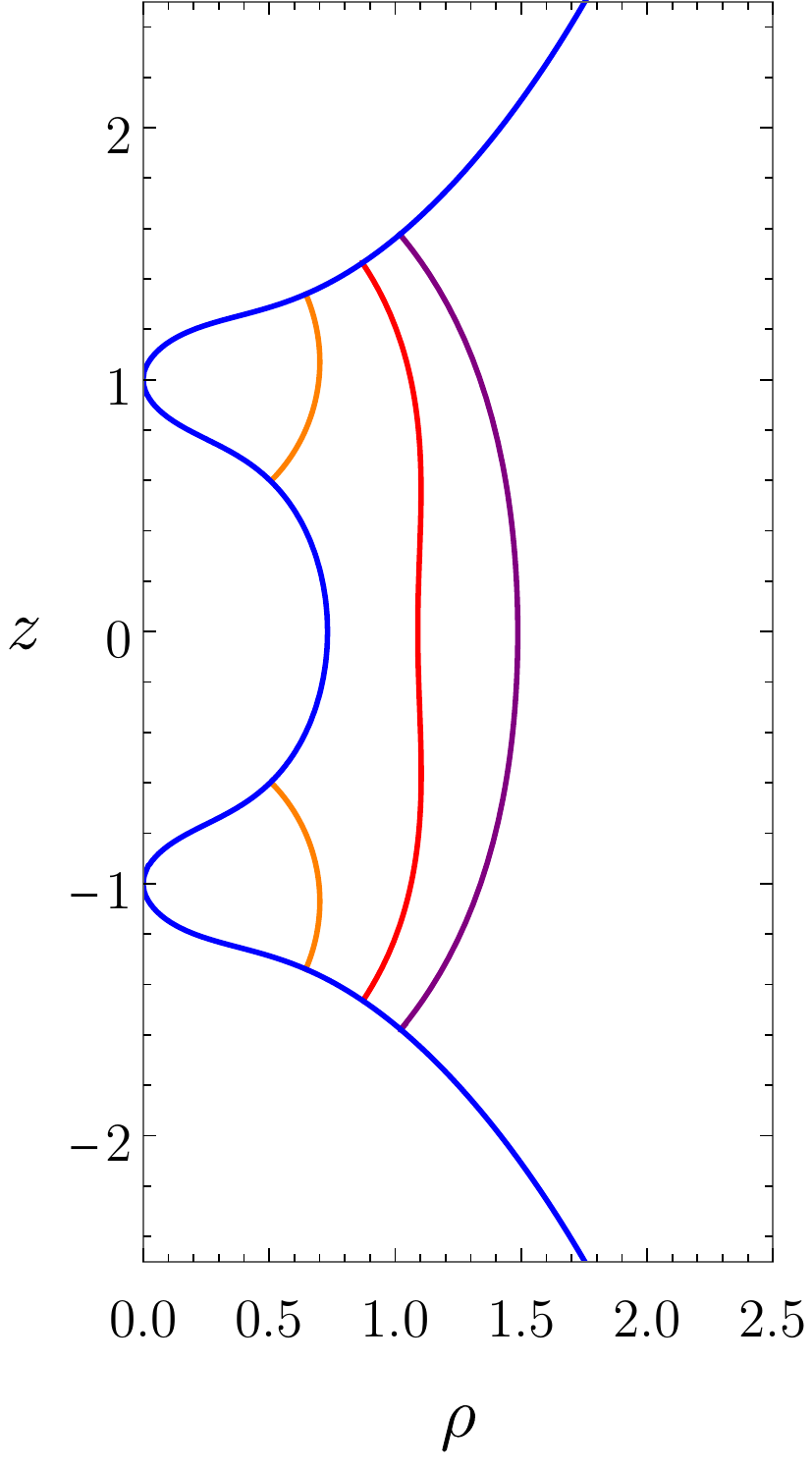} \label{fig:mp_fundamental_orbits_b}}
&
\subfigure[$p_{\phi}=5.1$]{\includegraphics[width=3.8cm]{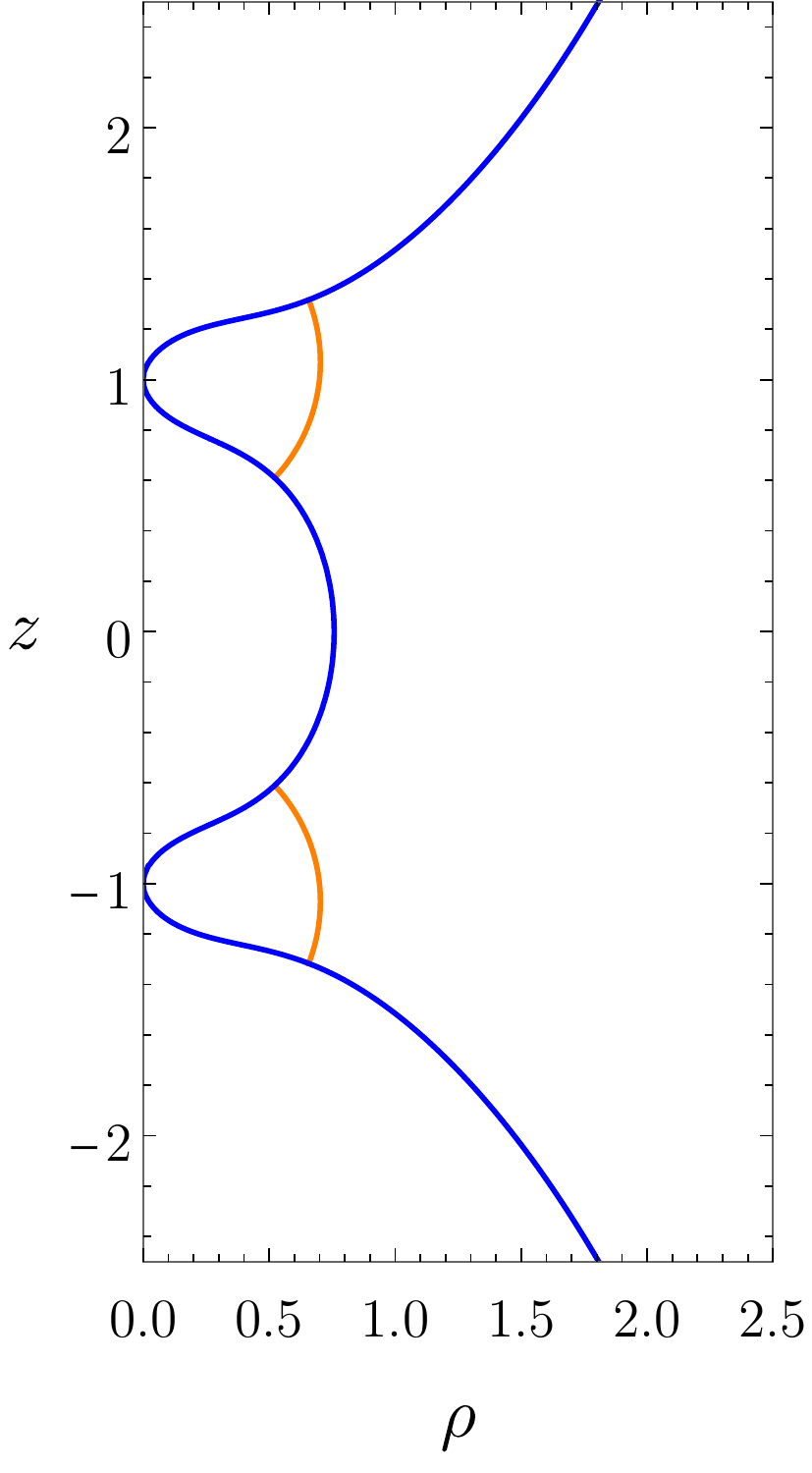} \label{fig:mp_fundamental_orbits_c}}
&
\subfigure[$p_{\phi}=5.92214$]{\includegraphics[width=3.8cm]{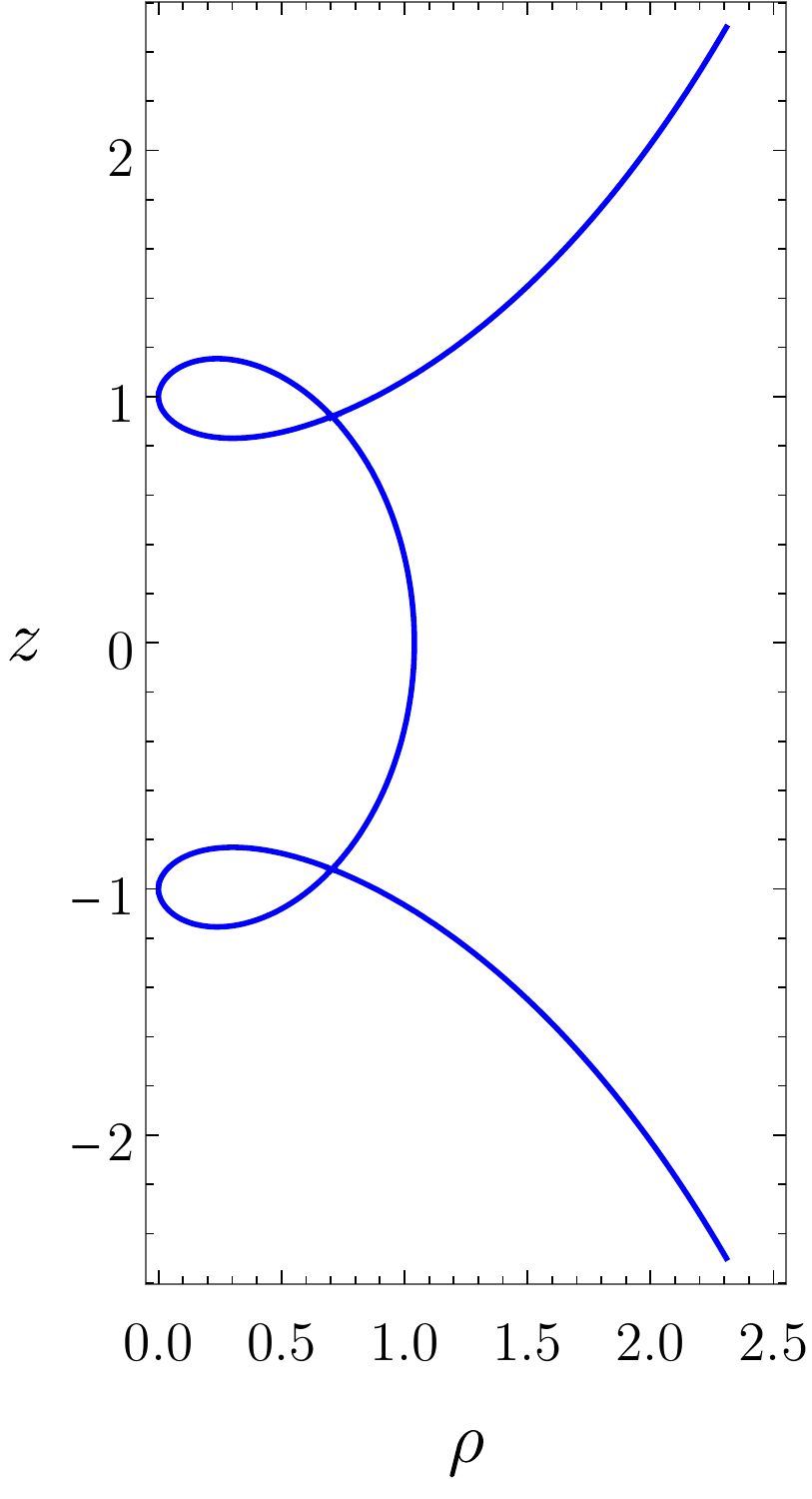} \label{fig:mp_fundamental_orbits_d}}
\end{tabular}
\caption{Examples of fundamental null orbits for an MP binary BH with coordinate separation $d = 2$. The two BHs appear as points, located at $(\rho, z) = (0, \pm 1)$. The blue curves are contours $h = p_{\phi}$. There are three types of fundamental null orbits in the $(\rho, z)$-plane: (I) a one-component light-ring [orange]; (II) a figure-of-eight orbit [red]; (III) a two-component light-ring [purple]. As one increases the value of $p_\phi$, orbits II and III move closer together, and then disappear. This corresponds to the end of the fractal region of the shadow. Orbit I exists up to $p_\phi \approx 5.92214$. This value corresponds to the end of the BH shadow, since absorption is forbidden by the contour $h = p_{\phi}$ for $p_\phi > 5.92214$.}
\label{fig:mp_fundamental_orbits}
\end{figure*}

\subsection{Photon orbits \label{sec:photon_orbits}}

In this section we describe the role played by a special class of photon orbits around the MP binary system: the \emph{fundamental photon orbits}. As described in Sec.~\ref{sec:hamiltonian}, the contours of a ``height function'' $h(\rho,z)$ [Eq.~\eqref{eqn:height_function}] demarcate the regions of phase space that are accessible to a ray with angular momentum $p_\phi$. For an equal-mass MP di-hole, a \emph{fundamental photon orbit} is a null geodesic $q^{\mu} (\lambda)$ with the following properties: (i) it is restricted to a compact subset of the $(\rho,z)$-plane; (ii) it is periodic, i.e., there is a value $T > 0$ such that $q^{\mu}(\lambda) = q^{\mu}(\lambda + T)$, for all $\lambda \in \mathbb{R}$; (iii) it is unstable; (iv) it touches the contour $h(\rho,z) = p_{\phi}$ in such a way that, locally, the ray is orthogonal to the contour; and (v) the radial momentum $p_{\rho}$ is zero where the orbit passes through the equatorial plane, by symmetry.

More general photon orbits are allowed in the MP di-hole system which satisfy some (but not all) of the above properties. For example, photon orbits which satisfy properties (i) and (ii), but which are \emph{stable} were explored in Ref.~\cite{DolanShipley2016}. A classification scheme for generic fundamental photon orbits in stationary axisymmetric spacetimes, which need only satisfy properties (i) and (ii) from the above list, is presented by Cunha \emph{et al.} in Ref.~\cite{CunhaHerdeiroRadu2017}. The role of light rings and fundamental photon orbits in the analysis of strong-field gravitational lensing is discussed in Ref.~\cite{CunhaHerdeiro2018}.

In Fig.~\ref{fig:mp_fundamental_orbits}, we show examples of the three types of fundamental photon orbit around the MP di-hole in the $(\rho, z)$-plane, labelled as follows: (I) a one-component light-ring; (II) a figure-of-eight orbit; (III) a two-component light-ring. Figures \ref{fig:mp_fundamental_orbits_a}--\ref{fig:mp_fundamental_orbits_d} show the effect on the fundamental orbits of changing the value of $p_{\phi}$, for the case $d=2$. As $p_{\phi}$ increases from zero, the contour $h(\rho,z) = p_{\phi}$ moves away from the symmetry axis, and orbits II and III move closer together. The orbits II and III merge at $p_{\phi} = \hat{p}_{\phi} \approx 5.08$. Type I orbits persist until the point where the contour ``pinches off'' at $p_\phi = p_{\phi}^{\ast} \approx 5.92214$.

If there exists a ray which passes asymptotically close to two or more fundamental orbits -- i.e., if fundamental orbits are ``dynamically connected'' -- then we anticipate that chaotic scattering phenomena will arise naturally \cite{Eckhardt:1987,Eckhardt:1988}. Indeed, it was demonstrated in Ref.~\cite{ShipleyDolan2016} that, for a given value of $p_\phi < p_{\phi}^{\ast}$, the 1D shadow is Cantor-like if the condition above is met (see e.g.~Figs.~7, 8 and 18(a) in \cite{ShipleyDolan2016}). However, it was also noted that it is not sufficient for two separate orbits of Type I to exist, because typically the inner orbits are not dynamically connected in the absence of the outer (Type II and III) orbits.

The 1D shadows for $d=2$ are observed to change in character as $p_\phi$ varies \cite{ShipleyDolan2016}. For $p_\phi > p_{\phi}^{\ast}$, the BHs are inaccessible, and the 1D shadow is the empty set. For $\hat{p}_{\phi} < p_{\phi} < p_{\phi}^{\ast}$, the outer orbits (Type II and III) do not exist, the inner orbits (Type I) are not dynamically connected, and the 1D shadow is regular, i.e., non-fractal. For $p_{\phi} < \hat{p}_{\phi}$, the inner orbits are dynamically connected with the outer orbits, and the 1D shadow has a Cantor-like fractal structure . In short, the appearance of fractal structure is directly linked to the existence of outer Type II/III orbits.

The 2D BH shadow is the union of 1D shadows. Thus, for $d=2$, the 2D shadow has parts which are regular and parts which are fractal. Mixed-modality shadows occur for coordinate separations $d$ such that the coexistence condition $\hat{p}_{\phi} < p_{\phi}^{\ast}$ is met. We show in the next section that the coexistence condition is only satisfied for sufficiently separated BHs with $d > \hat{d}$. For $d < \hat{d}$, the coexistence condition is not met, and thus we anticipate that the shadow will have no regular boundaries.

\subsection{The critical separation\label{sec:critical_separation}}

Here we describe a method to calculate the critical value $\hat{d}$ introduced in the previous section. We seek the  di-hole separation parameter $d$ which gives rise to a single outer fundamental orbit for $p_{\phi} = p_{\phi}^{\ast}$. That is, the value of $d$ for which the outer Type II and III orbits merge at exactly the value of $p_\phi$ at which the BHs become inaccessible.

First, we choose a value of $d$ and find the corresponding value of $p_{\phi}^{\ast}$ by using the method presented in Appendix B of Ref.~\cite{ShipleyDolan2016}. We then consider rays which start on the contour $h = p_{\phi}^{\ast}$ with $\rho = \rho_{0}$. The value of $z_{0} > 0$ is determined by numerically solving $h(\rho_{0}, z) = p_{\phi}^{\ast}$ for $z$. On the contour, $p_{\rho} = 0 = p_{z}$. We then evolve the geodesic equations for this choice of initial conditions until the ray passes through $z = 0$. At this point, we record the value of $\vartheta = \pi/2 + \arctan \left(p_{z}/p_{\rho}\right)$, where $\arctan \left(p_{z}/p_{\rho}\right)$ is the angle made by the tangent vector and the $\rho$-axis when the ray passes through the equatorial plane. By symmetry, the fundamental orbits II and III must have $\vartheta = 0$. Hence, the zeros of the function $\vartheta(\rho_{0})$ give the location of the fundamental orbits II and III.

\begin{figure}
\includegraphics[height=7cm]{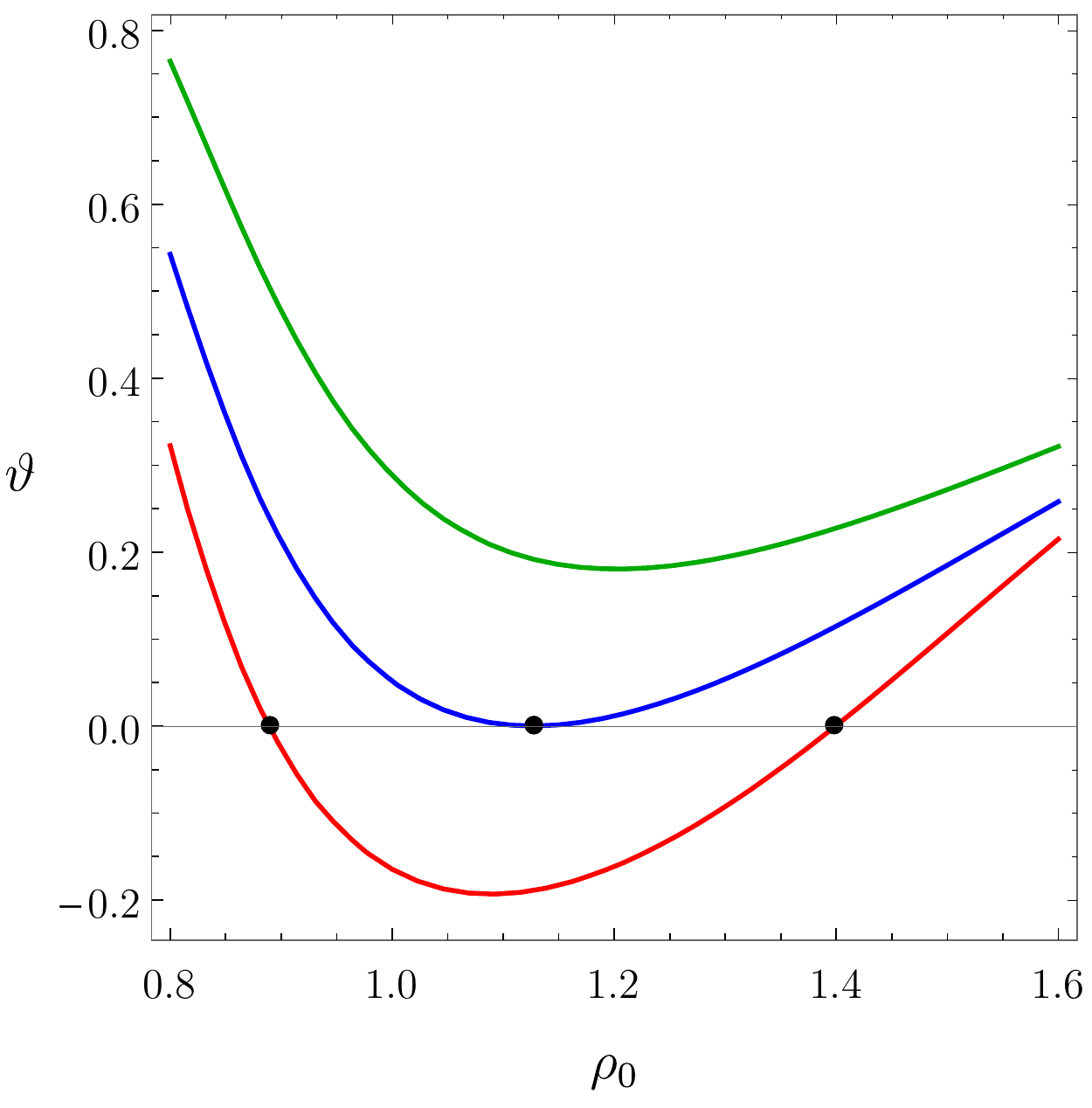}
\caption{Result of the algorithm used to search for fundamental orbits for three examples: (i) $d < \hat{d}$ [red]; (ii) $d = \hat{d}$ [blue]; (iii) $d > \hat{d}$ [green]. The function $\vartheta$ has (i) two zeros for $d < \hat{d}$; (ii) one zero for $d = \hat{d}$; and (iii) no zeros for $d > \hat{d}$. The zeros of $\vartheta$ are shown as black points in the figure.}
\label{fig:critical_d}
\end{figure}

Figure \ref{fig:critical_d} shows the function $\vartheta(\rho_{0})$ for three representative values of $d$. We seek the value of $d = \hat{d}$ for which $\vartheta(\rho_{0})$ admits a single zero, corresponding to the blue curve in Fig.~\ref{fig:critical_d}. We find that
\beq
\hat{d} \approx 1.2085 M,
\eeq
for the highly symmetric equal-mass MP di-hole. In the case of unequal masses, the $\mathbb{Z}_{2}$ reflection symmetry is broken. A more detailed analysis of the fundamental orbits would therefore be required to determine the value of $\hat{d}$ in this case.

For tightly bound di-holes with $d < \hat{d}$, there exist three types of fundamental orbits \emph{for all} $p_{\phi} \in [ 0, p_{\phi}^{\ast} ]$. In this regime, we anticipate that the MP di-hole shadow boundary will be entirely fractal. Conversely, for sufficiently separated di-holes with $d > \hat{d}$, there exists some $p_{\phi} = \hat{p}_{\phi}$, such that the outermost fundamental orbits no longer exist. We anticipate that the corresponding  shadows will have regular (i.e., non-fractal) parts.

\section{The method: Merging Wada basins \label{sec:wadamerge}}

The Wada property has its origins in topology. Three or more open sets are said to exhibit the Wada property if they share a common boundary \cite{yoneyama_theory_1917}. This counterintuitive situation appears naturally in nonlinear dynamical systems, where fractal geometry rules \cite{KennedyYorke1991}. Several methods have been proposed to test this striking property in dynamical systems; we briefly review these below.

Nusse and Yorke \cite{KennedyYorke1991,nusse_wada_1996} established that an unstable manifold crossing three (or more) basins could be used to prove the existence of Wada basin boundaries in phase space. However, this method cannot be applied in all circumstances, and it requires detailed knowledge of the system: an unstable trajectory starting on the boundary and crossing all of the basins must be found. This process can be cumbersome; indeed, many papers have been devoted to checking the Nusse--Yorke condition in a single dynamical system, for a particular set of parameters \cite{toroczkai_wada_1997, poon_wada_1996, Aguirre2001, aguirre_unpredictable_2002}. Later, a numerical method based on successive refinements of a grid was introduced. This approach allows one to test the Wada property in a variety of situations up to a given resolution \cite{DazaWagemakersSanjuanEtAl2015, daza_wada_2017}. Recently, a third numerical method has been proposed \cite{daza_ascertaining_2018}. This method involves merging the basins in a pairwise fashion, and comparing the boundaries of the merged basins with the original basins. Among the three methods outlined above, the merging method is the fastest and the only one able to provide a reliable test of the Wada property through simple examination of the basins at finite resolution, without computing new rays or invariant manifolds of the system.

\begin{figure}
\subfigure[Wada]{\includegraphics[height=7cm]{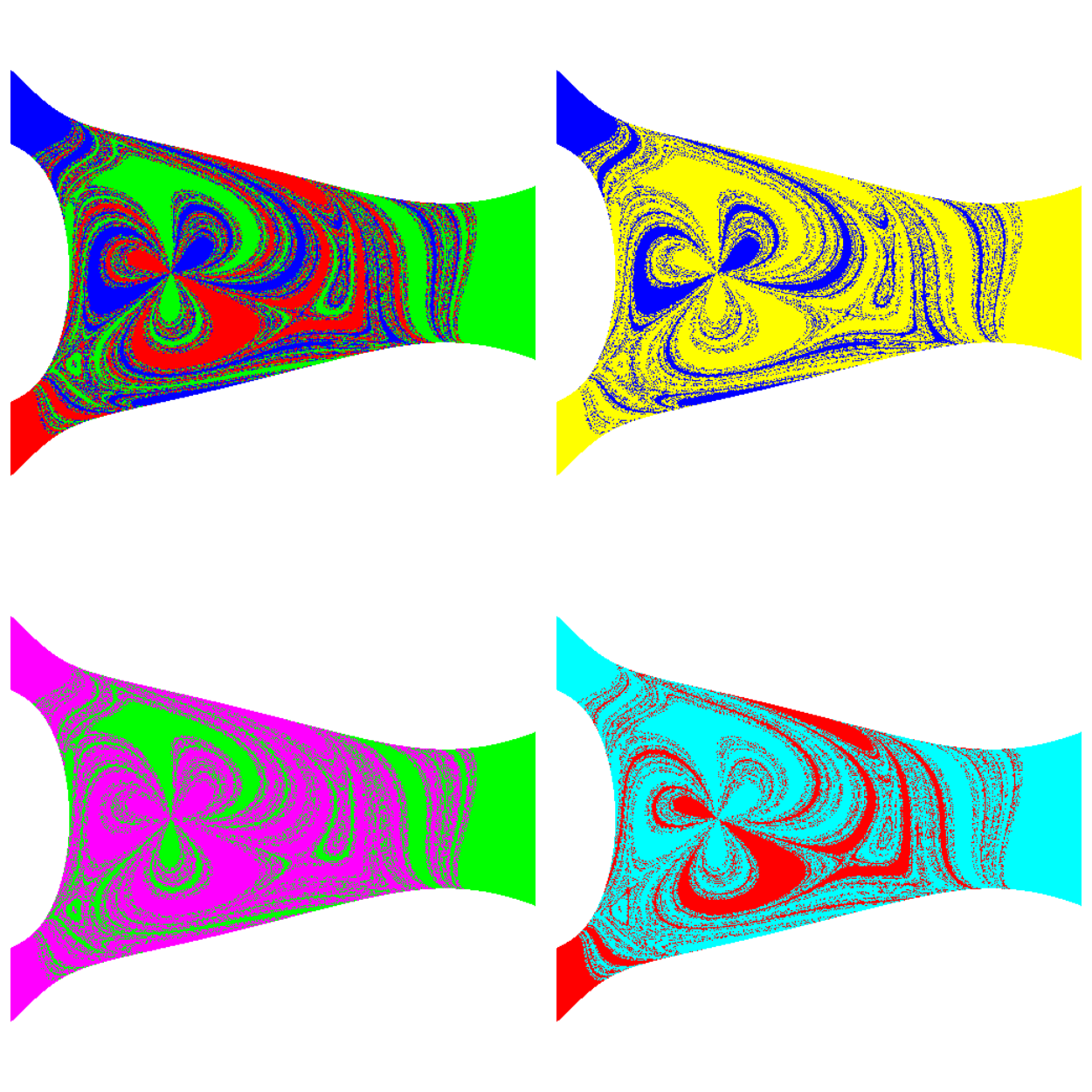} \label{fig:merging_basins_wada}}
\subfigure[Regular]{
\begin{tabular}{c}
\includegraphics[height=4cm]{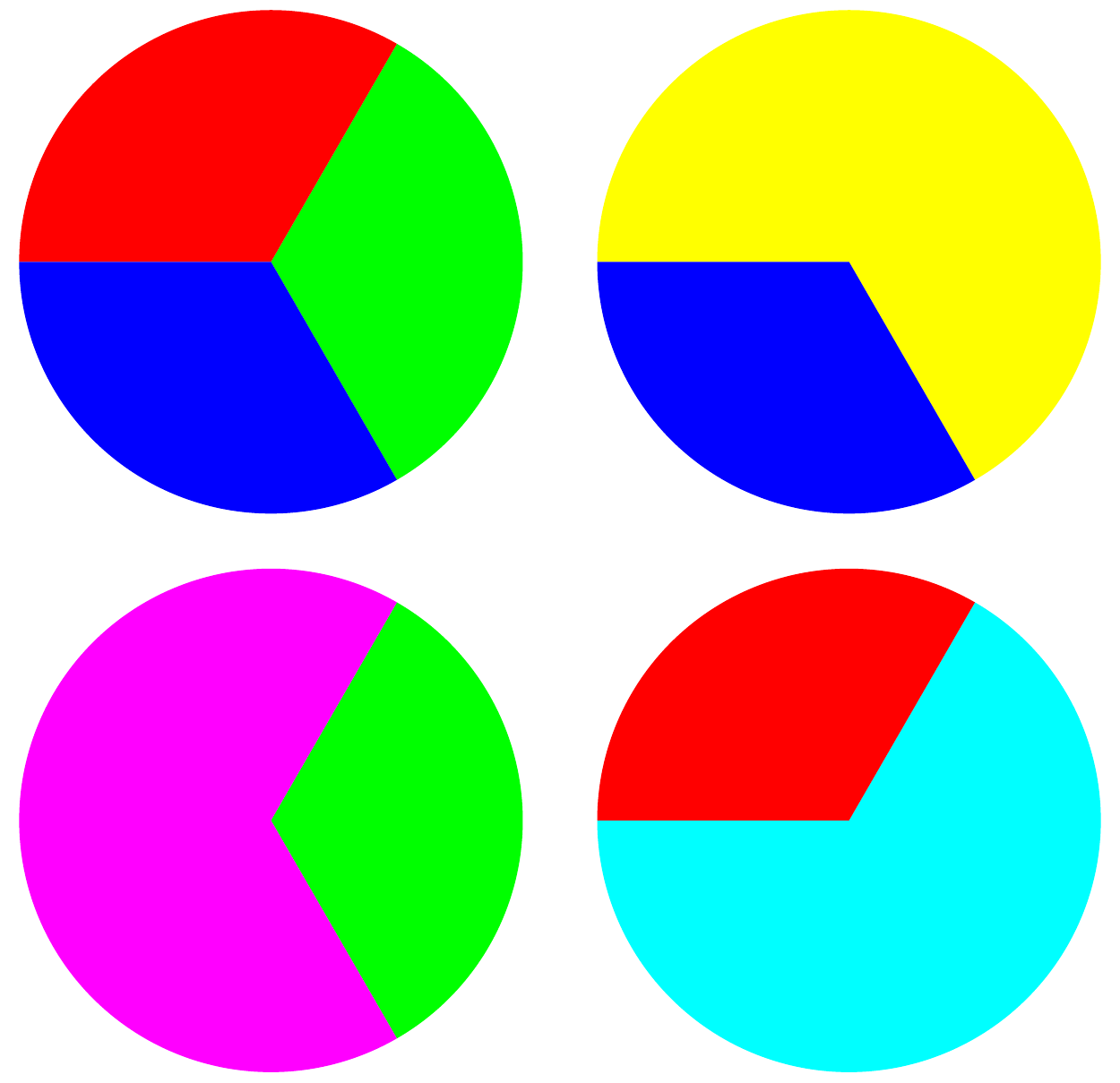}
\end{tabular}
\begin{tabular}{c}
\includegraphics[height=2cm]{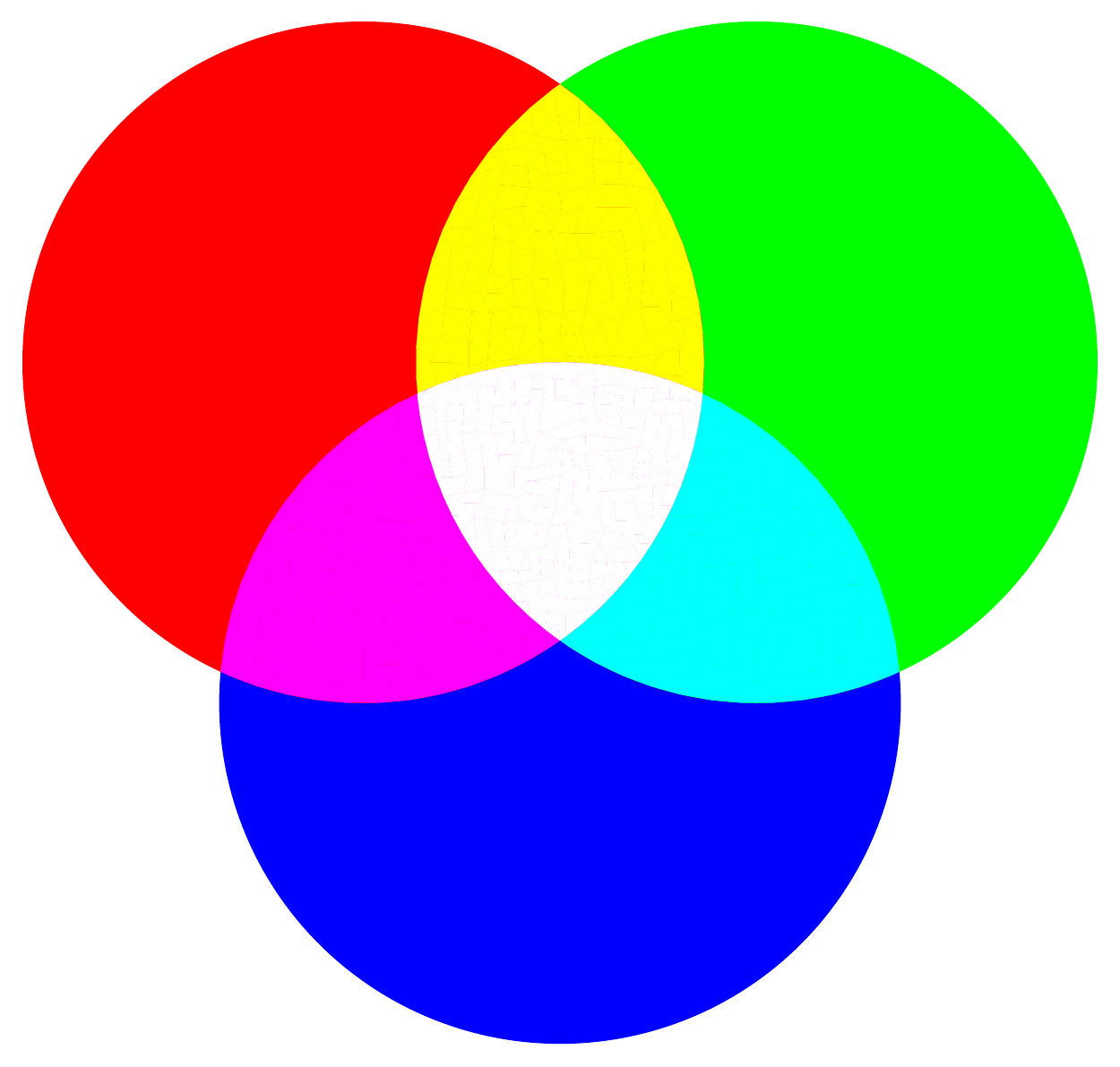}
\end{tabular}
 \label{fig:merging_basins_regular}
}
\caption{Illustration of the merging algorithm. (a) Merged Wada basins for the MP di-hole in $(\rho, z)$-space with $\Delta p_{\phi} = 0.03$. The top-left panel shows the original three-colour basins, as described in the text. The other three panels show the two-colour basins obtained from merging the original basins together. (b) A simple example of regular (non-fractal) basins and their merged versions. In this case, only one boundary point (the centre of the disc) is invariant under the pairwise merging of the basins; the remaining boundary points are not Wada points. The colour code is given in the right-hand plot. \label{fig:merging_basins}}
\end{figure}

The merging method is based upon the following counterintuitive observation: Wada boundaries are invariant under the action of merging any two of the basins together. In order to illustrate this property, we have depicted in Fig.~\ref{fig:merging_basins_wada} the exit basins of the MP di-hole system and their merged versions. At first glance, it may seem that we have simply changed the colours of the basins. However, a closer examination reveals that the boundaries are the same, but that in each case two of the three basins have been merged to form a new basin. Non-Wada boundaries change when the basins are merged, as shown in Fig.~\ref{fig:merging_basins_regular}. Using this feature we can test which basins are Wada based on which boundaries are invariant under the merging of the basins.

Given finite numerical resolution, it is impossible to ascertain whether two boundaries are exactly the same. In fact, the boundaries of the merged basins are slightly different even for Wada basins because of the finite grid of initial conditions used to realise the exit basin diagram. For this reason, we \textit{fatten} the boundaries by replacing each pixel by itself plus its $r$ nearest neighbours. The condition of the method states that if all the (original) \textit{slim boundaries} are contained in all the \textit{fat boundaries} then the basin possesses the Wada property. If this condition is fulfilled, we can say that the boundaries have the Wada property at a resolution determined by the internal parameter of the method $r$. The whole method relies on this fattening parameter $r$. We begin with $r = 1$ and increase its value until either the basins are classified as fully Wada or a stopping condition $r > r_{\mathrm{stop}}$ is reached. Of course, the merging method only ascertains that a basin is Wada up to a resolution determined by the fat pixels defined by the parameter $r$.

Here are the steps of the method:
\begin{enumerate}	
\item Begin with a finite resolution image of $N_{\mathrm{B}}$ exit basins. The method does not require any prior knowledge of the underlying dynamical system, but only the exit basins themselves.

\item Pick one basin, and merge all of the others to obtain a two-colour basin diagram. Repeat for each basin in turn. This yields $N_{\mathrm{B}}$ two-colour basin diagrams.

\item Identify the boundary in each two-colour basin diagram. This is achieved by identifying pixels with at least one neighbour of the opposite colour. This yields $N_{\mathrm{B}}$ \textit{slim boundaries}.

\item Fatten each slim boundary by a factor $r$ to obtain $N_{\mathrm{B}}$ \textit{fat boundaries}.

\item Take a fat boundary and test whether the union of slim boundaries is contained inside it. Repeat for each fat boundary in turn. If the union of slim boundaries is contained inside \emph{every} fat boundary, then the basins have the Wada property up to the resolution of the fat pixels. If this is not the case, then increase the value of $r$ and return to step 4, until the stopping condition $r > r_{\mathrm{stop}}$ is met. If the value $r_{\mathrm{stop}}$ is reached and  the union of slim boundaries is not contained in each fat boundary, then the method classifies the system as non-Wada. In the case of partial Wada boundaries \cite{zhang_wada_2013}, where Wada points and non-Wada points are present, the method can provide a list of the non-Wada points of the original image.
\end{enumerate}

\section{Results \label{sec:results}}

\subsection{The Wada property in phase space \label{sec:wadaexits}}

The exit basins of the HH Hamiltonian are known to exhibit the Wada property \cite{Aguirre2001}. Given its links with the MP di-hole system (see Sec.~\ref{sec:mpbinary} and Fig.~\ref{fig:equipotentials}), it is natural to speculate that the MP di-hole basins shown in Fig.~\ref{fig:mp_basins} will share this property \cite{ShipleyDolan2016}. Here we test this.

We applied the merging algorithm outlined in Sec.~\ref{sec:wadamerge} to the MP exit basins in both $(\rho, z)$-space and $(\rho, p_{\rho})$-space (see Fig.~\ref{fig:mp_basins}), using basin images with a resolution of $1000 \times 1000$ pixels. We tested every boundary point for the Wada property, for exit basis with $\Delta p_{\phi} = p_{\phi} - p_{\phi}^{\ast}$ in the range $\Delta p_{\phi} \in [0.02,0.15]$. For this choice of parameters, the merging algorithm classified \emph{all} boundary points as Wada points for fattening parameter $r = 3$.

As $\Delta p_{\phi} \rightarrow 0$, the width of the escapes also approaches zero, and KAM islands become dominant [see Figs.~\ref{fig:mp_basins_rho_z_1} and \ref{fig:mp_basins_rho_prho_1}], and the escape time for photons which start inside the scattering region blows up \cite{AguirreSanjuan2003}. It becomes computationally  expensive to verify the Wada property for small values of $\Delta p_{\phi}$. We have not verified the Wada property for widths $\Delta p_{\phi} < 0.02$; nevertheless, we expect all boundary points to remain Wada as $\Delta p_{\phi} \rightarrow 0$.

\subsection{The Wada property in black hole shadows \label{sec:wadashadows}}

We now examine the shadows of the MP di-hole, which are described in Sec.~\ref{sec:shadows} and shown in Fig.~\ref{fig:mpshadows}.

We applied the merging method (Sec.~\ref{sec:wadamerge}) to test for the Wada property in MP di-hole shadows, for various coordinate separations $d \in \left[0, 3\right]$ between the BHs. We generated the BH shadow images for an observer with a fixed viewing angle of $\theta = \pi/2$, by numerically integrating Hamilton's equations for a grid of $1000 \times 1000$ initial conditions (see Fig.~\ref{fig:mpshadows}).

\begin{figure}
\includegraphics[height=7cm]{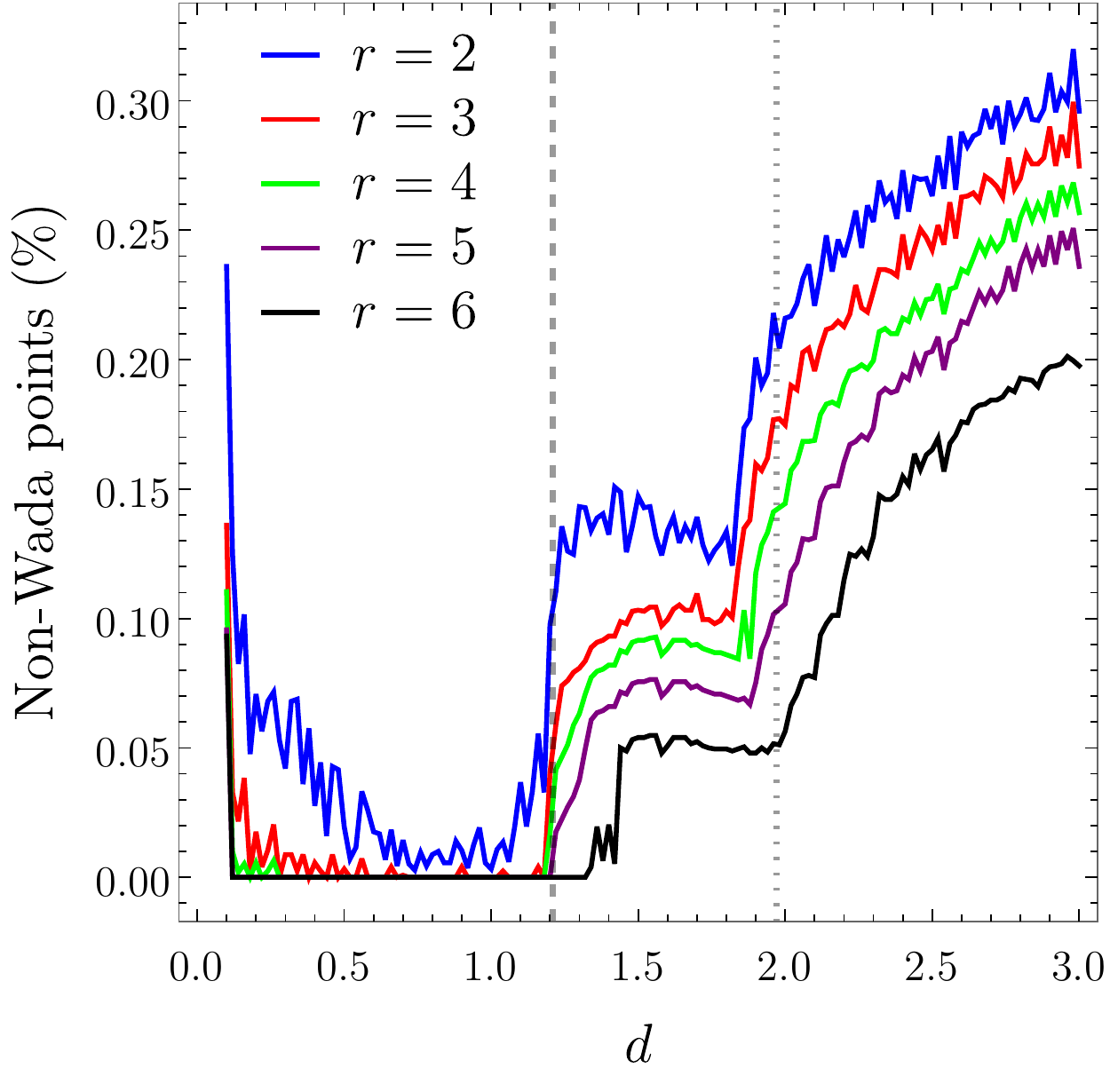}
\caption{Percentage of non-Wada points detected by the merging method for different values of the coordinate separation between the BHs $d$. The merging algorithm was performed for different values of the fattening parameter $r$. The dashed vertical line indicates the critical value $d = \hat{d} = 1.2085$, below which the shadows exhibit the full Wada property. The dotted vertical line, at $d = 1.97$, indicates the second jump in the number of non-Wada points. (See text for details.) \label{fig:nonwada_shadows}}
\end{figure}

The results of the algorithm are presented in Fig.~\ref{fig:nonwada_shadows}, which shows the percentage of boundary points which are \emph{not} classified as Wada points by the merging algorithm, as we vary the BH separation $d$. Figure \ref{fig:nonwada_shadows} provides evidence that the shadows are totally Wada (i.e., all boundary points are Wada points) for $0.1 \lesssim d \lesssim 1.2$. The algorithm suggests that there is a qualitative transition at $d \approx 1.2$, after which the shadow becomes only partially Wada.

A qualitative change of this kind was anticipated in our study of fundamental photon orbits in Secs.~\ref{sec:photon_orbits} and \ref{sec:critical_separation}. The existence of three types of fundamental periodic orbits (Fig.~\ref{fig:mp_fundamental_orbits}) for a fixed value of $p_{\phi}$ gives rise to Cantor-like structure in the 1D shadows of the MP di-hole. If all three types of fundamental orbits exist for $0 \le p_{\phi} < p_{\phi}^{\ast}$, then the 2D shadow will be totally Wada. Conversely, if there exists some value $\hat{p}_{\phi} < p_{\phi}^{\ast}$ for which the outer fundamental orbits cease to exist (e.g.~Fig.~\ref{fig:mp_fundamental_orbits_c}), then the 1D shadows with $\hat{p}_{\phi} < p_{\phi} < p_{\phi}^{\ast}$ will be regular, i.e., non-fractal. In such cases, the 2D shadow will be only partially Wada. We showed in Sec.~\ref{sec:critical_separation} that the latter is the case for $d > \hat{d} \approx 1.2085$. This value matches well with the observed transition in Fig.~\ref{fig:nonwada_shadows} (vertical line).

The results of the algorithm shown in Fig.~\ref{fig:nonwada_shadows} also suggest that there is a second qualitative change in the shadow structure at $d \approx 1.9$. It appears likely that the second transition occurs where the regular (i.e., non-fractal) region of the shadow touches the top of the main lobes of the shadow (see Fig.~\ref{fig:mpshadows}). For $d \gtrsim \hat{d}$, only the tips of the primary eyebrow-like features are regular. As one increases $d$, the regular region incorporates the top of the globular features in the centre of the shadow. Numerical investigation of the MP shadows indicates that this occurs at $d \approx 1.97$. This agrees well with the results of the Wada merge algorithm, and is shown as a dotted vertical line in Fig.~\ref{fig:nonwada_shadows}.

\begin{figure*}
\subfigure[]{\includegraphics[height=7cm]{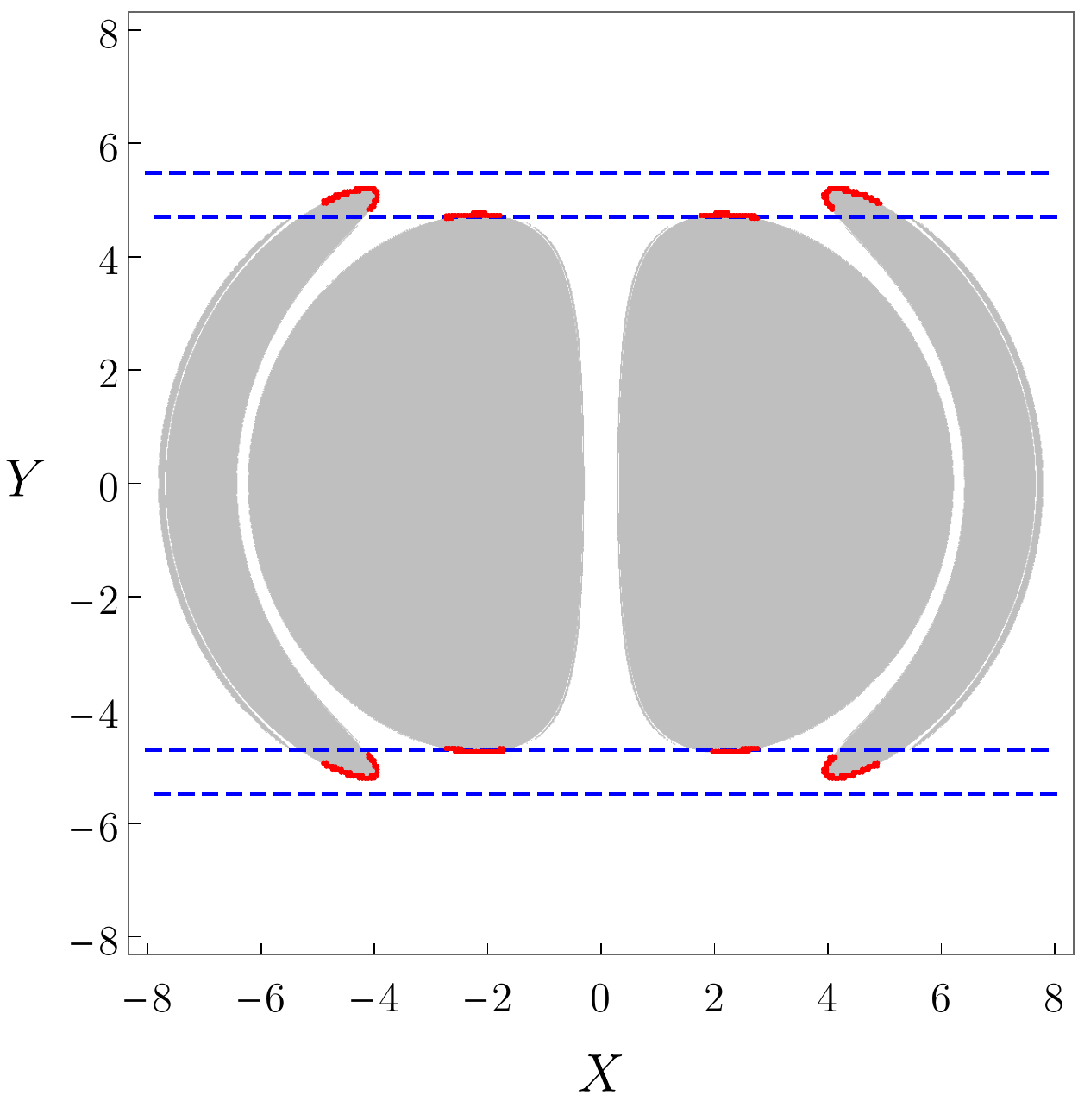} \label{fig:nonwada_points_a}} \hspace{1cm}
\subfigure[]{\includegraphics[height=7cm]{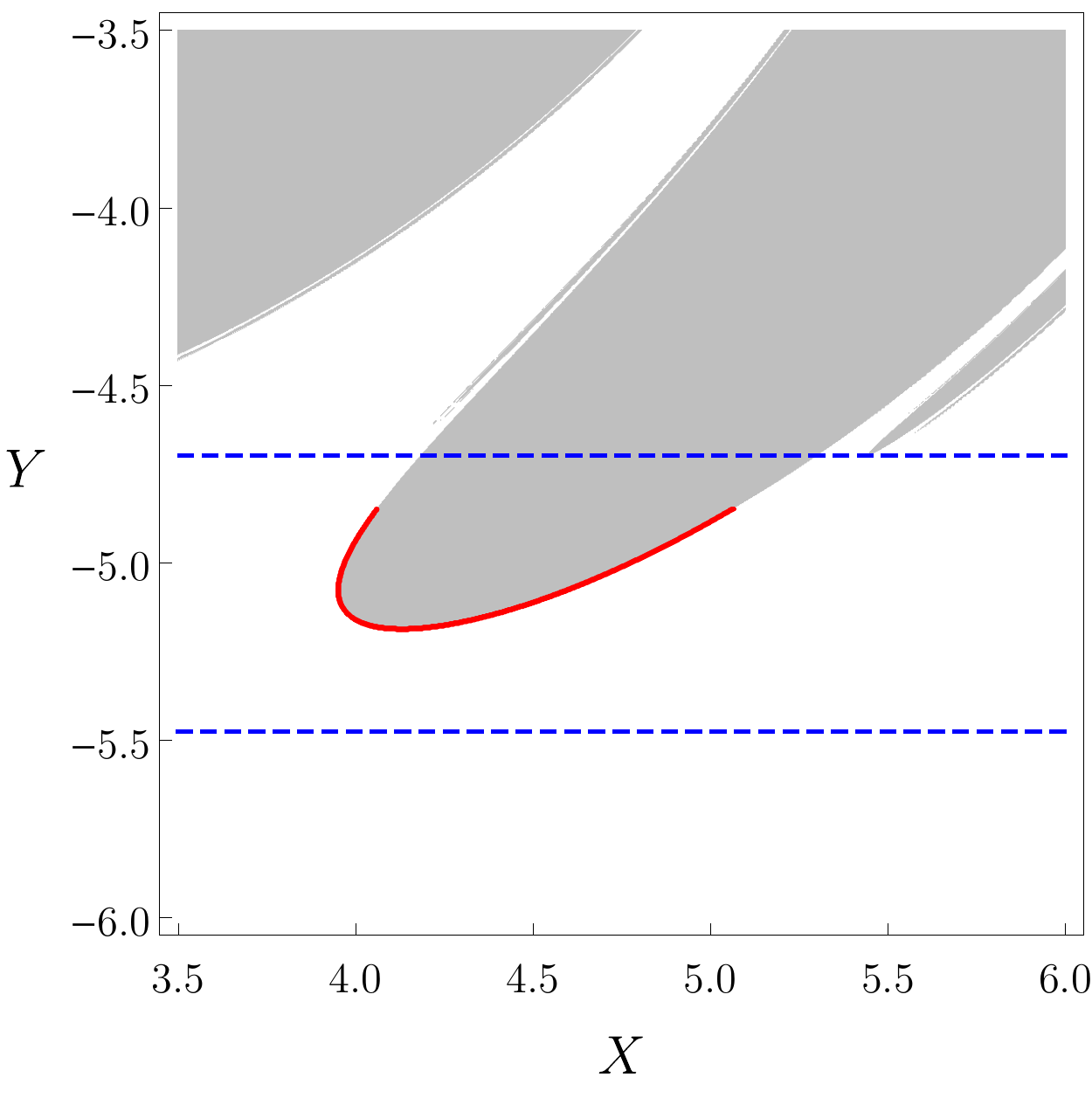} \label{fig:nonwada_points_b}}
\caption{(a) Shadow of the MP binary BH system for $d=2$. The shadows of the two BHs are shown in grey, whilst the basin corresponding to spatial infinity is plotted in white. The horizontal blue lines delimit the non-fractal regions, determined by looking at the critical values of $p_{\phi}$. The red points are the non-Wada points detected by the merging algorithm. (b) A magnified region of panel (a), which shows good agreement between the non-Wada points detected by the merging algorithm and the regular (i.e., non-fractal) regions of the BH shadow. \label{fig:nonwada_points}}
\end{figure*}

To confirm this interpretation, we used the merging method to highlight the non-Wada parts of the shadow. Figure \ref{fig:nonwada_points_a} shows the MP di-hole shadow for $d=2$ with an observer at a viewing angle $\theta = \pi/2$. The exit basins corresponding to the two BHs are both plotted in grey. The non-Wada points identified by the merging method are highlighted in red (with a fattening parameter $r=5$). The regular region, in which the shadow boundary is expected to be regular according to our analysis of fundamental orbits in Sec.~\ref{sec:photon_orbits}, lies between the horizontal blue dashed lines. The plot confirms that all the non-Wada points identified by the algorithm lie within that regular zone. Furthermore, it shows that for $d = 2$ the regular region has begun to impinge upon the main lobes.

In Fig.~\ref{fig:nonwada_points}, the agreement between the horizontal blue lines (determined by considering critical values of $p_{\phi}$) and the non-Wada points detected by the algorithm (red) could be improved by (i) increasing the resolution of the exit basin diagram, and (ii) taking $r_{\textrm{max}}$ (the location of the observer) to infinity. Both of these would make the algorithm more computationally expensive.

\section{Discussion and conclusions\label{sec:discussion}}

Here we have applied a new technique from nonlinear dynamics to study the fractal structures that arise in a binary BH model in general relativity. Remarkably, light rays on a MP di-hole are governed by a Hamiltonian dynamical system which has much in common with the H\'{e}non--Heiles system \cite{Aguirre2001}, i.e., the paradigmatic Hamiltonian for 2D time-independent chaotic scattering. We have analysed the dynamics of the MP di-hole -- modelling a pair of extremally charged BHs in static equilibrium -- in terms of exit basins in a plane (Fig.~\ref{fig:mp_basins}). We applied the Wada merge method \cite{daza_ascertaining_2018} (Sec.~\ref{sec:wadamerge}) to verify the Wada property in both (i) the exit basins in phase space (Sec.~\ref{sec:wadaexits}); and (ii) exit basins on an image plane which define the shadow cast by the BHs (Sec.~\ref{sec:wadashadows}).

We have demonstrated that the BH shadow can exhibit either the partial Wada or the full Wada property, depending on the value of the BH separation parameter $d$. The Wada property is typically associated with indeterminacy in a deterministic system. In this case, the final fate of a photon on a Wada boundary in phase space is uncertain, as it can end up in either of the BHs, or yet escape to spatial infinity.

Importantly, the algorithm of Ref.~\cite{daza_ascertaining_2018} does not use knowledge of the underlying dynamical system, or require computation of its invariant sets such as the unstable manifold. All that is needed as input is an exit basin image at finite resolution. A key result, shown in Fig.~\ref{fig:nonwada_points}, is that the algorithm successfully detected a ``phase transition'' in the BH shadow from fully Wada to partially Wada, at a certain value of the parameter $d$. This transition was anticipated from an analysis of fundamental photon orbits of the system (see Figs.~\ref{fig:critical_d} and \ref{fig:nonwada_points}). In cases where the underlying dynamical system is either unknown, or too complex to study analytically, the merging method offers a route to new physical insight.

The merging algorithm has several advantages over the Nusse--Yorke method \cite{nusse_wada_1996}. To verify that a basin is Wada, the Nusse--Yorke method requires the computation of an unstable manifold which crosses all of the exit basins in phase space. The image plane mixes phase space and parameter space: the coordinates on the observer's image plane are dependent on the phase space coordinates and the conserved parameter $p_{\phi}$. It is therefore unclear how one would construct an unstable manifold on the observer's image plane. Using the Nusse--Yorke method to test for the Wada property in BH shadows does not appear to be possible.

An open question is whether the shadows cast by BH binaries in Nature, such as the progenitor of GW150914, truly exhibit the Wada property. Sadly, although direct images of singleton shadows are anticipated soon \cite{BroderickJohannsenLoebEtAl2014}, there appears to be little prospect of direct imaging of binary shadows. However, realistic simulations from spectral codes in numerical relativity can now generate high-resolution 2D images of binary shadows \cite{BohnThroweHebertEtAl2015}. It would certainly be of interest to apply the merging method to classify high-resolution images as partially or fully Wada (or otherwise). Similarly, the method could be applied to shadows in other binary models \cite{CunhaHerdeiroRodriguez2018}.

To our knowledge, this work represents the first demonstration of the Wada property for a general-relativistic system. As well as demonstrating that tools from the field of chaos theory can be used to understand the rich dynamics of scattering processes in general relativity, this work highlights that there exist novel dynamical systems in gravitational physics which can be fruitfully explored by nonlinear dynamicists.

\section*{Acknowledgments}

This work has been supported by the Spanish State Research Agency (Agencia Estatal de Investigac\'{i}on) and the European Regional Development Fund (FEDER) under Project No.~FIS2016-76883-P. M.A.F.S.~acknowledges the jointly sponsored financial support by the Fulbright Program and the Spanish Ministry of Education (Program No.
FMECD-ST-2016). J.O.S.~acknowledges financial support from the University of Sheffield's Harry Worthington Scholarship. S.R.D.~acknowledges financial support from the Science and Technology Facilities Council (STFC) under Grant No.~ST/P000800/1, and from the European Union's Horizon 2020 research and innovation programme under the H2020-MSCA-RISE-2017 Grant No. FunFiCO-777740.

\bibliographystyle{apsrev4-1}
\bibliography{binary_black_hole_wada_refs}

\end{document}